\pdfoutput=1
\documentclass[manuscript]{acmart}
\usepackage{multirow}
\usepackage{verbatim}
\usepackage{rotating} 
\usepackage{tabularx}
\usepackage{longtable}
\usepackage{caption}
\usepackage{subfigure}
\newcommand{\bigcell}[2]{\begin{tabular}{@{}#1@{}}#2\end{tabular}}

\AtBeginDocument{%
  \providecommand\BibTeX{{%
    \normalfont B\kern-0.5em{\scshape i\kern-0.25em b}\kern-0.8em\TeX}}}

\setcopyright{acmcopyright}
\copyrightyear{2022}
\acmYear{2022}
\acmDOI{xxx.xxxx/xxxxx.xxxxxx}

\acmJournal{JACM}
\acmVolume{37}
\acmNumber{4}
\acmArticle{111}
\acmMonth{8}

\begin{document}
\title{A Survey on the Fairness of Recommender Systems}

\author{Yifan Wang}
\affiliation{%
  \institution{Tsinghua University}
  \city{Beijing}
  \country{China}
  \postcode{10084}
}
\email{yf-wang21@mails.tsinghua.edu.cn}

\author{Weizhi Ma}
\affiliation{%
  \institution{Institute for AI Industry Research (AIR), Tsinghua University}
  \city{Beijing}
  \country{China}
  \postcode{10084}
}
\email{mawz@tsinghua.edu.cn}

\author{Min Zhang$^*$}
\affiliation{%
  \institution{Tsinghua University}
  \city{Beijing}
  \country{China}
  \postcode{10084}
}
\email{z-m@tsinghua.edu.cn}

\author{Yiqun Liu}
\affiliation{%
  \institution{Tsinghua University}
  \city{Beijing}
  \country{China}
  \postcode{10084}
}
\email{yiqunliu@tsinghua.edu.cn}

\author{Shaoping Ma}
\affiliation{%
  \institution{Tsinghua University}
  \city{Beijing}
  \country{China}
  \postcode{10084}
}
\email{msp@tsinghua.edu.cn}

\def\authors{Yifan Wang, Weizhi Ma, Min Zhang, Yiqun Liu, and Shaoping Ma}

\thanks{
$*$ Corresponding author.\\
This work is supported by the Natural Science Foundation of China (Grant No. U21B2026, 62002191) and Tsinghua University Guoqiang Research Institute.\\
Authors' addresses: Yifan Wang, Min Zhang, Yiqun Liu, and Shaoping Ma, Department of Computer Science and Technology, Institute for Artificial Intelligence, Beijing National Research Center for Information Science and Technology, Tsinghua University, Beijing 100084, China. Emails: yf-wang21@mails.tsinghua.edu.cn, z-m@tsinghua.edu.cn, yiqunliu@tsinghua.edu.cn, msp@tsinghua.edu.cn; Weizhi Ma, Institute for AI Industry Research (AIR), Tsinghua University, Beijing 100084, China. Email: mawz@tsinghua.edu.cn. 
}

\renewcommand{\shortauthors}{Wang, et al.}

\begin{abstract}
Recommender systems are an essential tool to relieve the information overload challenge and play an important role in people’s daily lives. Since recommendations involve allocations of social resources (e.g., job recommendation), an important issue is whether recommendations are fair. Unfair recommendations are not only unethical but also harm the long-term interests of the recommender system itself. As a result, fairness issues in recommender systems have recently attracted increasing attention.
However, due to multiple complex resource allocation processes and various fairness definitions, the research on fairness in recommendation is scattered. To fill this gap, we review over 60 papers published in top conferences/journals, including TOIS, SIGIR, and WWW. 
First, we summarize fairness definitions in the recommendation and provide several views to classify fairness issues.
Then, we review recommendation datasets and measurements in fairness studies and provide an elaborate taxonomy of fairness methods in the recommendation.  
Finally, we conclude this survey by outlining some promising future directions.

\end{abstract}




\maketitle

\section{Introduction}
Nowadays, the amount of information available on the Internet has far exceeded individuals' information needs and processing capacity, which is known as information overload \cite{200037}. As a tool to alleviate information overload, recommender systems are widely used in people's daily lives (e.g., news recommendations, career recommendations, and even medical recommendations) and play a crucial role. Utility (such as click-through rate, dwell time, etc.) has been the most vital metric for recommender systems. However, only considering utility may lead to problems like the Matthew effect \cite{200035} and filter bubble \cite{200036}. Hence more views of recommender system performance have been proposed, such as diversity, efficiency, privacy, etc. Fairness is one of these critical issues. Recommender systems serve a resource allocation role in society by allocating information to users and exposure to items. Whether the allocation is fair can affect the personal experience and social good \cite{201907}.

Fairness problems have received increasing attention from academia, industry, and society. Unfairness exists in different recommendation scenarios and various resources for both users and items. 
For users, there are significant differences in the recommendation accuracy between users of different ages and genders in movie recommendations and music recommendations, with female users and older users getting worse recommendation results \cite{201813}. In addition to accuracy, existing studies have also found considerable differences in other recommendation measurements such as diversity and novelty \cite{202119}. 
For items, existing research has found that minority items could get worse ranking performance and less exposure opportunity \cite{201906,202017}. Besides, in premium business scenarios, paid items may receive worse services from the platform than non-paid items \cite{202201}.
Moreover, there are potential unfairness issues with various recommendation methods. Both traditional recommendation methods \cite{201813} and deep learning models \cite{202107} can suffer from unfairness.

Mitigating these unfairness phenomena is of great importance for recommender systems. Here are some reasons.
(1) from an \textbf{ethical perspective}, as early as ancient Greece, fairness was listed by Aristotle as one of the crucial virtues to make people live well \cite{200038}. Fairness is an important virtue, and a fundamental requirement for a just society \cite{200039}.
(2) from a \textbf{legal perspective}, Anti-discrimination laws \cite{200034} require that employment, admissions, housing, and public services do not discriminate against different groups of people based on gender, age, race, etc. For example, minority-owned companies should be recommended at a similar rate to white-owned companies in a job recommendation scenario \cite{202011}.
(3) from a \textbf{user perspective}, a fair recommender system facilitates the exposure of different information in the recommendations, including some niche information, which may help break the information cocoon, alleviate the societal polarization, broaden users' horizons and enhance the value of recommendations.
(4) from an \textbf{item perspective}, a fair recommender system can allocate more exposure to long-tail items, alleviating the Matthew effect \cite{202112}. It may also motivate these providers of niche items and then improve the diversity and creativity of items.
(5) from a \textbf{system perspective}, a fair recommender system is conducive to its long-term interest. For example, an unfair recommender system may recommend popular content for users with niche interests, resulting in a bad experience. Similarly, it may also provide little exposure for niche providers. The lack of positive feedback may lead to a tendency for niche groups to leave the platform, which will reduce the diversity of content and users on the platform in the long run and affect the platform's growth \cite{202001}.
Therefore, addressing unfairness is a critical issue for recommender systems. 

A concept closely related to fairness is bias, which has also attracted extensive attention in current years. Some biases in recommender systems can lead to unfairness problems, such as popularity bias \cite{202104} and mainstream bias \cite{202105}. There are also some biases that have little to do with fairness, such as position bias. Generally speaking, fairness reflects normative ideas about how a recommender system should be, while bias is more concerned with statistical issues, such as the difference between what the model learns and the real world.

Although fairness has been studied in computer science for decades \cite{200025}, and there is a lot of related work in machine learning \cite{200016,200018}, fairness in recommendation has its unique problems. 
First, recommender systems are two-sided platforms that serve both users and items, where two-sided fairness needs to be guaranteed.
Second, fairness in recommendation is dynamic in nature as there exists a feedback loop between users and the system.
Third, on most platforms, the recommendation needs to be personalized by considering the unique needs of each user. Fairness in recommendation should also take users' personalization into account.
Furthermore, apart from accuracy, fairness needs to be jointly considered with other measurements in the recommendation, such as diversity, explainability, and novelty.
Therefore, current fairness work in machine learning, which mainly focuses on classification, could hardly be leveraged in recommender systems directly.

\begin{figure}[h]
  \centering
  \includegraphics[width=0.6\linewidth]{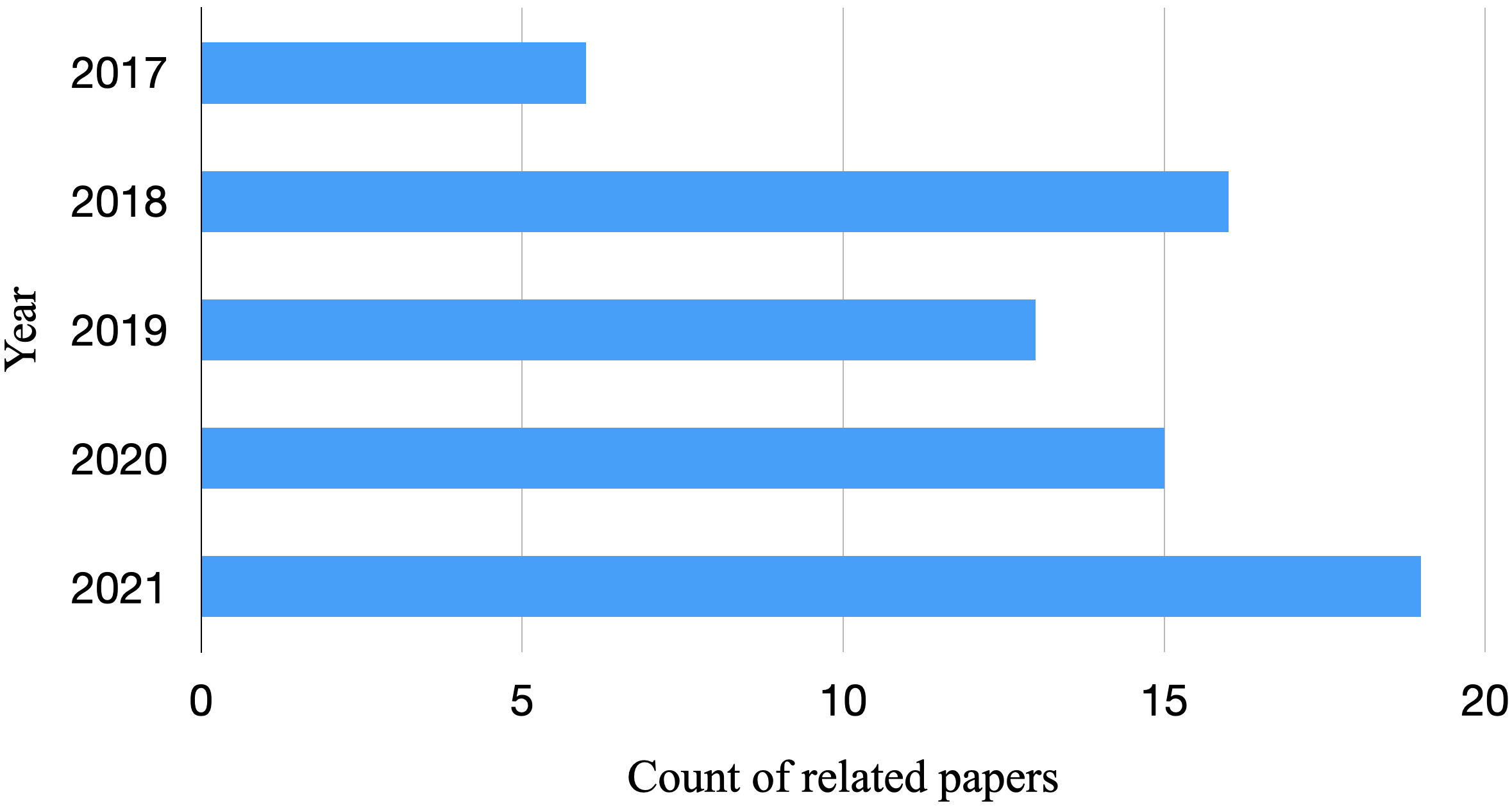}
  \caption{The statistics of publications related to fairness in recommendation. We omit the work in 2022 in the figure, as most of the work in 2022 has not been published.}
  \Description{The number of the publications related to fairness in recommendation in 2017, 2018, 2019, 2020, and 2021 are 6, 16, 13, 15, and 18, respectively.]}
  \label{years}
\end{figure}

For the above reasons, fairness in recommendation has become an important topic in the research community. The attracted attention is increasing, which trends have been shown in Fig.\ref{years}. As shown in Table \ref{maintable}, more than sixty fairness-related papers about recommendations have been published in top IR-related conferences and journals (e.g., TOIS, SIGIR, WWW, and KDD) in recent five years. In the table, researches on fairness are summarized with their different definitions, targets, subjects, granularity, and optimization objects (details on these definitions are given in Table \ref{tab:def} and Section 3). We can find the focus of current studies. For example, consistent fairness (CO) is the most common definition of fairness, and current studies mainly focus on the group level. These trends are further discussed in the corresponding sections below.

{
\footnotesize
\begin{longtable}{l|l|cc|ccc|cc|cc|l|l}
\caption{A lookup table for the reviewed papers about fairness in recommendation (Here "CO" means consistent fairness, "CA" means calibrated fairness, "CF" means counterfactual fairness, "EF" means envy-free fairness, "RMF" means Rawlsian maximin fairness, "PR" means process fairness and "MSF" means maximin-shared fairness, details on these definitions are given in Table \ref{tab:def}).} \label{maintable} \\
\hline
\endfirsthead
\bottomrule
\endfoot
\caption{(continued)} \\
\toprule\toprule
\multirow{2}{*}{\textbf{Paper}} & \multirow{2}{*}{\textbf{Def.}} & \multicolumn{2}{c|}{\textbf{Target}} &  \multicolumn{3}{c|}{\textbf{Subject}} & \multicolumn{2}{c|}{\textbf{Granularity}} & \multicolumn{2}{c|}{\textbf{Optim. Object}} & \multirow{2}{*}{\textbf{Pub.}} & \multirow{2}{*}{\textbf{Year}} \\[0.1em]
\cline{3-4}\cline{5-7}\cline{8-9}\cline{10-11}
\rule{0em}{2.5ex}
                       &                             & Group & Individual         & User & Item & Joint         & Single & Amortized & Treatment & Impact & & \\
\midrule
\endhead
\toprule\toprule
\multirow{2}{*}{\textbf{Paper}} & \multirow{2}{*}{\textbf{Def.}} & \multicolumn{2}{c|}{\textbf{Target}} &  \multicolumn{3}{c|}{\textbf{Subject}} & \multicolumn{2}{c|}{\textbf{Granularity}} & \multicolumn{2}{c|}{\textbf{Optim. Object}} & \multirow{2}{*}{\textbf{Pub.}} & \multirow{2}{*}{\textbf{Year}} \\[0.1em]
\cline{3-4}\cline{5-7}\cline{8-9}\cline{10-11}
\rule{0em}{2.5ex}
                       &                             & Group & Individual         & User & Item & Joint         & Single & Amortized & Treatment & Impact & & \\
\midrule
\midrule
\cite{202204}          &   CA                        & \checkmark &           &          & \checkmark &            & &  \checkmark    &  \checkmark       &  & WSDM & 2022 \\   
\hline
\cite{202205}          &   CA                        & \checkmark &           &          & \checkmark &            & \checkmark &     &  \checkmark       &  & WSDM & 2022 \\   
\hline
\cite{202206}          &   EF                        &  & \checkmark          &   \checkmark       &  &            & \checkmark &     &  \checkmark       &  & WSDM & 2022 \\   
\hline
\cite{202203}          &   CO                        &  &  \checkmark         &          & \checkmark &            & &  \checkmark    &  \checkmark       &  & WWW & 2022 \\   
\hline
\cite{202202}          &   EF                        &  & \checkmark           &   \checkmark         & &            & \checkmark &     &         & \checkmark  & AAAI & 2022 \\   
\hline
\cite{202201}          &   CA                        & \checkmark &            &   \checkmark         & \checkmark &            &  &    \checkmark     &         & \checkmark  & BIGDATARES. & 2022 \\   
\hline
\cite{202121}          &   CO                        & \checkmark &            &            & \checkmark &            &  &    \checkmark     &    \checkmark     &   & TOIS & 2021 \\   
\hline
\cite{202122}          &   RMF                        &  & \checkmark           &            &  &  \checkmark          &  &    \checkmark     &    \checkmark     &   & NIPS & 2021 \\   
\hline
\cite{202101}          & CO \& CA                           & \checkmark &            &            & \checkmark &            & \checkmark &         &    \checkmark     & \checkmark   & WSDM & 2021 \\   
\hline
\cite{202102}          & CO                          & \checkmark &            &              &   \checkmark &            &              & \checkmark &   \checkmark      &              & WSDM                         & 2021                  \\ \hline
\cite{202103}          &  CO                           &   \checkmark &            &              &   \checkmark &            &   \checkmark &            &   \checkmark      &              & WSDM                         & 2021                  \\ \hline
\cite{202105}          &  CO                           &   \checkmark &            &   \checkmark &              &            &              & \checkmark &                   & \checkmark   & WSDM                         & 2021                  \\ \hline
\cite{202106}          &  CO                           &   \checkmark &            &   \checkmark &              &            &              & \checkmark &   \checkmark      &              & AAAI                         & 2021                  \\ \hline
\cite{202107}          &  CO                           &   \checkmark &            &   \checkmark &              &            &              & \checkmark &                   & \checkmark   & WWW                          & 2021                  \\ \hline
\cite{202108}          & CO \& CA  &   \checkmark &            &              &   \checkmark &            &   \checkmark &            &   \checkmark      &              & WWW                          & 2021                  \\ \hline
\cite{202109}          & PR                              &   -          & -          &   \checkmark &              &            &   -          & -          &   -               & -            & WWW                          & 2021                  \\
\hline
\cite{202110}          &  CA                           &   \checkmark &            &              &   \checkmark &            &              & \checkmark &   \checkmark      &              & WWW                          & 2021                  \\ \hline
\cite{202112}          &  CO                           &   \checkmark &            &   \checkmark &              &            &              & \checkmark &                   & \checkmark   & WWW                          & 2021                  \\ \hline
\cite{202114}          &  CO \& CA   &   \checkmark & \checkmark &              &              & \checkmark &              & \checkmark &   \checkmark      &              & SIGIR                        & 2021                  \\ \hline
\cite{202115}          & CF                       &              & \checkmark &   \checkmark &              &            &   \checkmark &            &   \checkmark      &              & SIGIR                        & 2021                  \\ \hline
\cite{202116}          &  CA                           &   \checkmark &            &              &   \checkmark &            &              & \checkmark &   \checkmark      & \checkmark   & SIGIR                        & 2021                  \\ \hline
\cite{202117}          & RMF                     &              & \checkmark &              &   \checkmark &            &              & \checkmark &   \checkmark      &              & SIGIR                        & 2021                  \\ \hline
\cite{202118}          & RMF                     &   \checkmark & \checkmark &              &   \checkmark &            &   \checkmark &            &   \checkmark      &              & KDD                          & 2021                  \\ \hline
\cite{202119}          &  CO                           &   \checkmark &            &   \checkmark &              &            &              & \checkmark &                   & \checkmark   & RECSYS                       & 2021                  \\ \hline
\cite{202120}          &  CO                           &              & \checkmark &              &   \checkmark &            &              & \checkmark &                   &              & CIKM                         & 2021                  \\  \hline
\cite{202004}          &  CO                           &              & \checkmark &              &   \checkmark &            &              & \checkmark &   \checkmark      &              & UMAP                         & 2020                  \\ \hline
\cite{202005}          &  CO                           &   \checkmark &            &              &   \checkmark &            &   \checkmark &            &   \checkmark      &              & UMAP                         & 2020                  \\
\hline
\cite{202013}          &  CO                           &   \checkmark &            &   \checkmark &              &            &              & \checkmark &   \checkmark      &              & UMAP                         & 2020                  \\
\hline
\cite{202006}          & CO \& CA   &   \checkmark & \checkmark &              &   \checkmark &            &   \checkmark &            &   \checkmark      &              & CIKM                         & 2020                  \\ \hline
\cite{202008}          & CO \& CA   &   \checkmark &            &              &              & \checkmark &              & \checkmark &   \checkmark      & \checkmark   & WSDM                         & 2020                  \\ \hline
\cite{202009}          & EF \& MSF &              & \checkmark &              &              & \checkmark &              & \checkmark &   \checkmark      &              & WWW                          & 2020                  \\ 
\hline
\cite{202015}          &  CO                           &   \checkmark &            &              &   \checkmark &            &   \checkmark &            &   \checkmark      &              & WWW                          & 2020                  \\
\hline
\cite{202010}          &  CO                           &              & \checkmark &   \checkmark &              &            &   \checkmark &            &   \checkmark      &              & RECSYS                       & 2020                  \\ \hline
\cite{202011}          &  CA                           &   \checkmark &            &              &   \checkmark &            &              & \checkmark &   \checkmark      &              & PAKDD                        & 2020                  \\ \hline
\cite{202012}          &  CO                           &              & \checkmark &              &   \checkmark &            &              & \checkmark &   \checkmark      &              & ACCESS                  & 2020                  \\ 
\hline
\cite{202003}          &  CO                           &   \checkmark & \checkmark &   \checkmark &              &            &              & \checkmark &   \checkmark      &              & SIGIR                        & 2020                  \\
\hline
\cite{202014}          &  CA                           &   \checkmark &            &              &   \checkmark &            &              & \checkmark &   \checkmark      & \checkmark   & SIGIR                        & 2020                  \\
\hline
\cite{202017}          &  CO                           &   \checkmark &            &              &   \checkmark &            &              & \checkmark &   \checkmark      & \checkmark   & SIGIR                        & 2020                  \\
\hline
\cite{202016}          &  CO                           &   \checkmark &            &              &   \checkmark &            &   \checkmark &            &                   & \checkmark   & AAAI                         & 2020                  \\  \hline
\cite{202018}          &  CO                           &              & \checkmark &   \checkmark &              &            &   \checkmark &            &   \checkmark      &              & SAC                          & 2020                  \\ \hline
\cite{201901}          &  CA                           &   \checkmark &            &   \checkmark &   \checkmark &            &              & \checkmark &   \checkmark      & \checkmark   & RMSE              & 2019                  \\ \hline
\cite{201902}          &  CA                           &   \checkmark &            &              &   \checkmark &            &   \checkmark &            &   \checkmark      &              & RMSE              & 2019                  \\ \hline
\cite{201904}          & PR                              &   -          & -          &   \checkmark &   \checkmark &            &   -          & -          &   -               & -            & ICML                         & 2019                  \\ \hline
\cite{201905}          &  CA                           &   \checkmark &            &              &   \checkmark &            &   \checkmark &            &   \checkmark      &              & KDD                          & 2019                  \\ \hline
\cite{201906}          &  CO                           &   \checkmark &            &              &   \checkmark &            &   \checkmark &            &                   & \checkmark   & KDD                          & 2019                  \\ \hline
\cite{201907}          &  CO                           &   \checkmark & \checkmark &   \checkmark &   \checkmark &            &              & \checkmark &                   & \checkmark   & WSDM                         & 2019                  \\ \hline
\cite{201908}          &  CO                           &   \checkmark &            &              &   \checkmark &            &   \checkmark &            &   \checkmark      &              & RECSYS                       & 2019                  \\ \hline
\cite{201909}          &  CO                           &   \checkmark &            &   \checkmark &              &            &              & \checkmark &   \checkmark      &              & UMAP                         & 2019                  \\ \hline
\cite{201910}          & CO \& CA   &   \checkmark &            &   \checkmark &              &            &              & \checkmark &   \checkmark      &              & LOCALREC                     & 2019                  \\ \hline
\cite{201912}          &  CO                           &              & \checkmark &              &   \checkmark &            &              & \checkmark &   \checkmark      &              & MEDES                        & 2019                  \\ \hline
\cite{201913}          &  CO                           &              & \checkmark &   \checkmark &              &            &   \checkmark &            &   \checkmark      &              & SAC                          & 2019                  \\ \hline
\cite{201916}          &  CA                           &              & \checkmark &              &   \checkmark &            &   \checkmark &            &   \checkmark      &              & NIPS                         & 2019                  \\ \hline
\cite{201801}          &  CO                           &   \checkmark &            &   \checkmark &              &            &              & \checkmark &                   & \checkmark   & FATREC              & 2018                  \\ \hline
\cite{201802}          & -                                             &   \checkmark &            &   \checkmark &              &            &   -          & -          &   -               & -            & FATREC              & 2018                  \\ \hline
\cite{201815}          &  CA                           &   \checkmark &            &              &   \checkmark &            &              & \checkmark &   \checkmark      &              & RECSYS                       & 2018                  \\ \hline
\cite{201816}          &  CA                           &   \checkmark &            &              &   \checkmark &            &   \checkmark &            &   \checkmark      &              & RECSYS                       & 2018                  \\ \hline
\cite{201817}          &  CA                           &   \checkmark &            &              &   \checkmark &            &              & \checkmark &   \checkmark      &              & RECSYS                       & 2018                  \\ \hline
\cite{201805}          &  CO                           &   \checkmark &            &              &   \checkmark &            &   \checkmark &            &   \checkmark      &              & UMAP                         & 2018                  \\ \hline
\cite{201806}          &  CO                           &              & \checkmark &   \checkmark &              &            &              & \checkmark &   \checkmark      &              & WWW                          & 2018                  \\
\hline
\cite{201820}          &  CO                           &   \checkmark &            &              &   \checkmark &            &              & \checkmark &   \checkmark      &              & WWW                          & 2018                  \\
\hline
\cite{201808}          &  CA                           &              & \checkmark &              &   \checkmark &            &              & \checkmark &   \checkmark      &              & SIGIR                        & 2018                  \\ \hline
\cite{201810}          &  CO                           &   \checkmark &            &              &   \checkmark &            &   \checkmark &            &   \checkmark      &              & CIKM                         & 2018                  \\ \hline
\cite{201811}          &  CO                           &   \checkmark &            &   \checkmark &   \checkmark &            &              & \checkmark &   \checkmark      &              & CIKM                         & 2018                  \\ \hline
\cite{201809}          &  CO                           &   \checkmark &            &   \checkmark &   \checkmark &            &              & \checkmark &   \checkmark      &              & FAT*                         & 2018                  \\ \hline
\cite{201813}          &  CO                           &   \checkmark &            &   \checkmark &              &            &              & \checkmark &                   & \checkmark   & FAT*                         & 2018                  \\ \hline
\cite{201814}          &  CO                           &   \checkmark &            &   \checkmark &   \checkmark &            &              & \checkmark &                   & \checkmark   & FAT*                         & 2018                  \\ \hline
\cite{201818}          &  CA                           &   \checkmark &            &              &   \checkmark &            &   \checkmark &            &   \checkmark      &              & ICALP                        & 2018                  \\ \hline
\cite{201819}          &  CA                           &   \checkmark &            &              &   \checkmark &            &   \checkmark &            &   \checkmark      &              & KDD                          & 2018                  \\  \hline
\cite{201702}          &  CO                           &   \checkmark &            &   \checkmark &              &            &              & \checkmark &                   & \checkmark   & NIPS                         & 2017                  \\ \hline
\cite{201703}          & EF                             &              & \checkmark &   \checkmark &              &            &   \checkmark &            &   \checkmark      &              & WWW                          & 2017                  \\ \hline
\cite{201704}          &  CO                           &              & \checkmark &   \checkmark &              &            &   \checkmark &            &   \checkmark      &              & RECSYS                       & 2017                  \\ \hline
\cite{201705}          &  CA                           &   \checkmark &            &              &   \checkmark &            &              & \checkmark &   \checkmark      &              & PAKDD                        & 2017                  \\ \hline
\cite{201707}          &  CO                           &   \checkmark &            &              &   \checkmark &            &   \checkmark &            &   \checkmark      &              & CIKM                         & 2017                  \\ \hline
\cite{201708}          &  CO                           &   \checkmark &            &              &   \checkmark &            &   \checkmark &            &   \checkmark      &              & SSDBM                        & 2017                  \\
\bottomrule
\end{longtable}
}

Research on the fairness in recommendation is blossoming. However, due to various scenarios, diverse stakeholders, and different measurements, the research on fairness in the recommendation field is scattered. In order to fill this gap, this survey systematically reviews the existing formally published research on fairness in the recommendation from several perspectives. The corresponding summary and discussion can guide and inspire future work. In summary, the contributions of this survey are as follows.

\begin{itemize}
    \item We summarize existing \textbf{definitions} of fairness in recommendation and provide several \textbf{views} for classifying fairness issues in recommendation.
    \item We introduce some widely used \textbf{measurements} for fairness in recommendation and review fairness-related recommendation \textbf{datasets} in previous studies.
    \item We review current \textbf{methods} for fair recommendations and provide an elaborate taxonomy of methods.
    \item We outline several promising \textbf{future research directions} from the perspective of definition, evaluation, algorithm design, and explanation.
\end{itemize}

Several surveys are related to the topic of this survey. As far as we know, Castillo \cite{201911} firstly reviews fairness and transparency in information retrieval briefly. However, it only covers the related work before 2018, and in recent years, fairness in recommendations has developed greatly. \cite{200026,200049} concentrate on fairness in machine learning, but fairness in recommendation is not covered, especially its unique characteristics.
Chen et al. \cite{200027} recently reviewed bias in recommender systems and introduced fairness issues, but fairness is not their main focus, and fairness measurements and datasets are not covered.
To the best of our knowledge, there is no survey dedicated to systemically reviewing and detailing the fairness in the recommendation in a complete view. 

This survey is structured as follows. In Section 2, we introduce existing definitions of fairness in the recommendation and discuss some related concepts. In Section 3, we present several perspectives to classify fairness issues in the recommendation. In Section 4, we introduce representative measurements for measuring fairness in the recommendation. In Section 5, we provide a taxonomy of methods to address unfairness in the recommendation. In Section 6, we introduce fairness-related datasets in recommender systems. In Section 7, we present possible future research directions. Finally, we conclude this survey in Section 8.

\section{Definitions of Fairness in Recommendation}
\begin{figure}[h]
  \centering
  \includegraphics[width=0.83\linewidth]{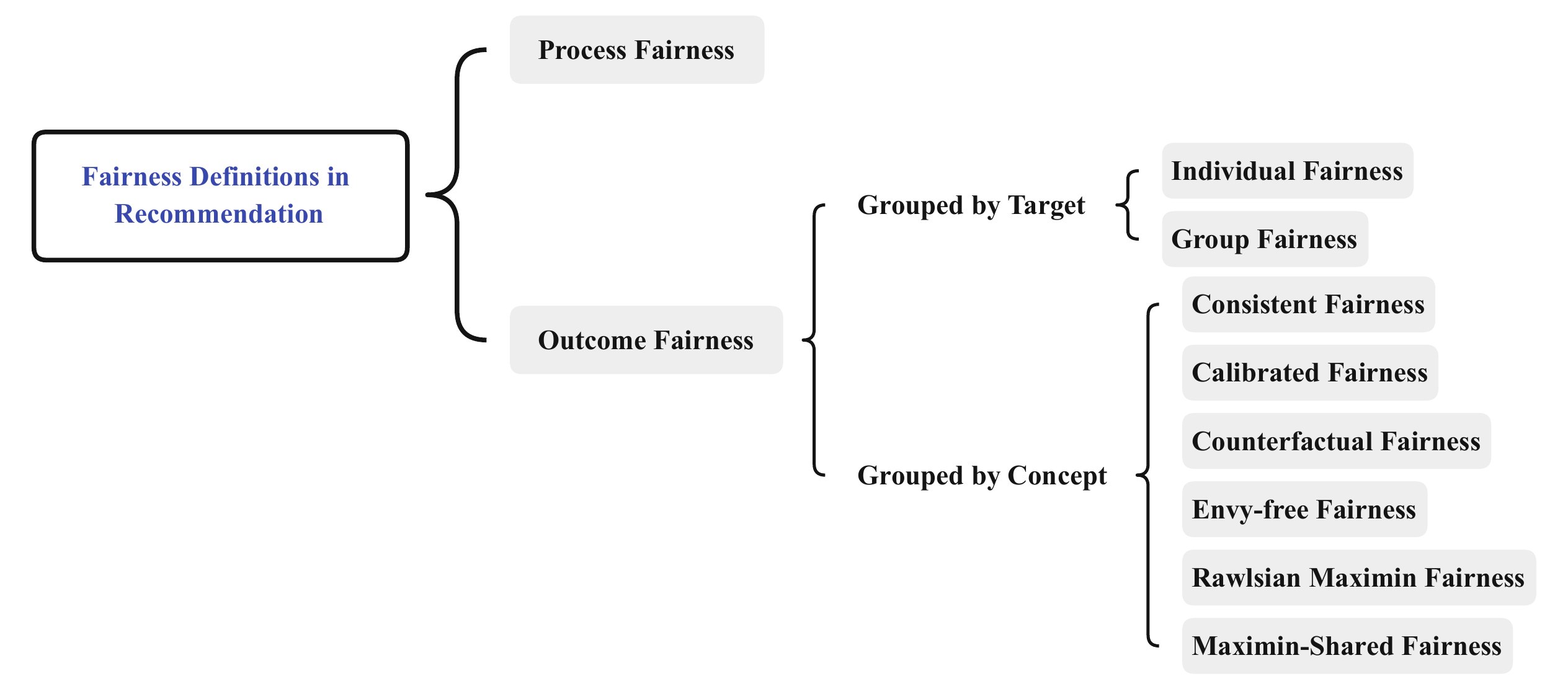}
  \caption{Taxonomy of fairness definitions in recommendation.}
  \Description{The figure is fully described in the text.}
  \label{deffig}
\end{figure}

In this section, we first provide definitions of fairness and then discuss the relationship between fairness and some related concepts in recommender systems. It is worth noting that discussions about fairness have existed since ancient times, but there is still no consensus on fairness. Due to the multitude of discussions related to fairness, it is impossible to list all relevant definitions. Therefore, we will introduce several definitions of fairness appearing in the research on recommendation, which can also be applied to other domains. The taxonomy of the reviewed fairness definitions is illustrated in Fig.\ref{deffig}. To our knowledge, the definitions listed here are sufficient to cover the research on fairness in the recommendation. Besides, the notations used in the definitions are shown in Table \ref{tab:def_notations}.

\subsection{Fairness Definitions}

\begin{table}[h]
\caption{Notations used in fairness definitions and their explanations}
\label{tab:def_notations}
\begin{tabular}{l|l}
\toprule \toprule
\textbf{Notation} & \textbf{Explanation}              \\
\hline
$i$     & the i-th individual (e.g., a user or an item)                        \\
$G_i$      & the i-th group of individuals                     \\
$H$        & the hypothesis space of recommendation models              \\
$C_h(\cdot)$     & the outcome (e.g., predicted scores or recommendation lists) of model $h$ given individuals or groups                                 \\
$D(\cdot,\cdot)$       & the distance function between individuals or groups\\
$D_o(\cdot,\cdot)$       & the distance function between outcomes\\
$V(\cdot)$          &  the value function of outcomes     \\
$V_i(\cdot)$          &  the personalized value function of  outcomes for certain individual $i$      \\
$M(\cdot)$    & the merit function of individuals or groups                      \\
\bottomrule \bottomrule
\end{tabular}
\end{table}

As we mentioned in the introduction, recommender systems play a resource allocation role in society, allocating information to users and exposure to items. For allocation, there are two aspects worthy of attention. One is the allocation process, such as the fairness of the recommendation model. The other is the allocation outcome, such as the fairness of the information received by users.
Depending on whether the focus is on the process or outcome, fairness can be divided into process fairness and outcome fairness.
\subsubsection{Process Fairness}

\begin{itemize}
    \item \textit{\textbf{Process Fairness.} Process fairness believes that the fair allocation should be fair in process \cite{200028,200003}, which is also called procedural justice \cite{200003}.}
\end{itemize}

Existing studies \cite{200003,200005,200061} generally focus on whether the information utilized in the allocation process is fair. In the case of job recommendation, process fairness concerns whether the recommendation model is fair, such as whether some unfair features (e.g., race) are used and whether the learned representations are fair.

\subsubsection{Outcome Fairness}

\begin{itemize}
    \item \textit{\textbf{Outcome Fairness.} Outcome fairness holds that the fair allocation should lead to fair outcomes \cite{200017,200003}, which is also called distributive justice \cite{200003}.}
\end{itemize}

For example, in the case of job recommendation, outcome fairness concerns the recommendation outcome, such as whether whites would be more likely to be recommended than blacks even if they have the same ability.

The difference between these two kinds of fairness is similar to the difference between teleology and deontology in ethics. Teleology believes that whether a behavior is good or bad is related to outcomes, while deontology believes that it is only related to processes \cite{200040}.

As the majority of existing research in recommendation focuses on the fairness of outcomes, we concentrate on definitions related to outcome fairness in the following. Outcome fairness can be further sub-grouped according to the target and concept.

\smallskip
\noindent \textbf{Grouped by Target}.
Based on whether the target is to ensure group-level or individual-level fairness, outcome fairness can be further categorized into \textit{group fairness} and \textit{individual fairness}.

\begin{itemize}
\item  \textit{\textbf{Group Fairness.} Group fairness holds that outcomes should be fair among different groups. }
\end{itemize}

There are various ways to divide groups, and the most common is based on some explicit fairness-related attributes, such as gender, age, and race. When there are multiple fairness-related attributes, the whole may be divided into numerous subgroups. Fairness should be considered in these subgroups, as even if the groups under each single attribute division are fair, subgroups may be unfair to each other \cite{200032,200033}.

\begin{itemize}
\item \textit{\textbf{Individual Fairness.} Individual fairness believes that outcomes should be fair at the individual level.}
\end{itemize}

Individual fairness in some work refers to the idea that similar individuals should be treated similarly \cite{201808,200012}. However, there are also other definitions of individual-level fairness. For the sake of clarity, we use \textit{individual fairness} to refer to a more general definition, i.e., fairness at the individual level.

Group fairness is more complex than individual fairness as different divisions may exist, and the divisions may be dynamic, i.e., one individual may belong to different groups at different times \cite{202102}. 
Moreover, individual fairness can be theoretically regarded as a special case of group fairness, in which each individual belongs to a unique group. 

\smallskip
\noindent \textbf{Grouped by Concept.} Although fairness can be classified according to the target, we do not know what kind of outcomes are fair up to this point. About this, different researchers have different opinions, which we call fairness concepts. These concepts reflect researchers' understanding of what requirements should be met for fair outcomes. Compared to targets, fairness concepts include more information about fairness, and we can give more concrete formal definitions. We present these fairness concepts in the following.

A lot of papers \cite{201808,202102,202112} define fairness based on the similarity of the input (i.e., the individuals or groups receiving the allocation) and the output (i.e., the outcome of the allocation), which we call \textit{consistent fairness}.
\begin{itemize}
\item \textit{\textbf{Consistent Fairness.} At the individual level, consistent fairness argues that similar individuals should be treated similarly \cite{201808}. Formally, a fair model $h$ should satisfy: for any two individuals $i$ and $j$, if $D(i,j) \approx 0$, then $D_o(C_h(i),C_h(j)) \approx 0$. At the group level, consistent fairness requires that different groups should be treated similarly \cite{202102,202112}. Formally, a fair model $h$ should satisfy: for any two groups $G_i$ and $G_j$, it has $D_o(C_h(G_i),C_h(G_j)) \approx 0$.}
\end{itemize}

This concept of fairness first appeared in Aristotle's quote, "like cases should be treated alike" \cite{200038}, which is thought to describe the consistency of fairness \cite{200029}. Dwork et al. \cite{200012} first formalized this definition at the individual level using the Lipschitz condition in a classification task. As a proper distance function of individuals is difficult to define, existing studies \cite{201907, 202003} in recommendation usually use a trivial case as an alternative of consistent fairness, in which all individuals (or groups) are assigned similar outcomes. For the distance function of outcomes, current work often uses the difference between specific metrics (e.g., NDCG for users \cite{201813}) to measure distance.

\begin{itemize}
    \item \textit{\textbf{Calibrated Fairness.}
Calibrated fairness \cite{201816} requires that the value of the outcome of an individual (or group) should be proportional to its merit, which is also called merit-based fairness \cite{202014}. Formally, at the individual level, a fair model $h$ should satisfy: for any two individuals $i$ and $j$, it has $\frac{V(C_h(i))}{M(i)} = \frac{V(C_h(j))}{M(j)}$. The group-level formalization is similar and only requires replacing $i,j$ with $G_i, G_j$.}
\end{itemize}

This concept of fairness is closely related to Adams' Equity Theory \cite{200041}. Calibrated fairness requires two functions to measure the merit of individuals (or groups) and the value of the allocation outcome. The measure of merit often depends on the scenario, and the measure of value usually are some commonly used metrics (e.g., CTR for items \cite{202014}).

\smallskip
Most research focuses on the above two concepts of fairness, while a small number of papers explore other concepts.

\begin{itemize}
    \item \textit{\textbf{Envy-free Fairness.}
Envy-free fairness requires that individuals should be free of envy, i.e., they should not be jealous of the results of others' outcomes \cite{200042, 202202}. Formally, a fair model $h$ should satisfy: for every individual $i$ and its outcome $C_h(i)$, it has $V_i(C_h(i)) \ge V_i(C_h(j))$ for any other individual $j$.}
\item \textit{\textbf{Counterfactual Fairness.}
Counterfactual fairness requires that individuals have the same outcome in the real world as they do in the counterfactual world \cite{200043}. This means that if an individual belongs to a different group from the current one, its outcome will not change. Formally, a fair model $h$ should satisfy: for every individual $i$, it has $C_h(i) = C_h(i)_{i \in G_j}$ for any other group $G_j$. The counterfactual $C_h(i)_{i \in G_j}$ can be calculated according to Pearl's three steps \cite{200043}.}
\item \textit{\textbf{Rawlsian Maximin Fairness.}
Rawlsian maximin fairness requires maximizing the value of the outcome of the worst individual or group \cite{200039}. Formally, at the individual level, a fair model $h$ should satisfy: $h^* = \mathop{argmax}\limits_{h \in H} \min\limits_i V(C_h(i))$. The group-level formalization is similar and only requires replacing $i$ with $G_i$.}
\item \textit{\textbf{Maximin-shared Fairness.}
Maximin-shared fairness requires all individuals (or groups) to receive better outcomes than their maximin share \cite{200042}. Formally, a fair model $h$ should satisfy: for every individual $i$ and its outcome $C_h(i)$, given its personal value function $V_i(\cdot)$, we should have $V_i(C_h(i)) \ge MMS_i$, here $MMS_i = \max\limits_{h \in H} \min\limits_{j} V_i(C_h(j))$. The group-level formalization is similar and only requires replacing $i,j$ with $G_i, G_j$.}
\end{itemize}

\smallskip
Table \ref{tab:def} divides the reviewed papers according to their definitions. It can be found that existing research pays more attention to outcome fairness. While in outcome fairness, previous studies mainly concentrate on group fairness in terms of target and focus on consistent fairness and calibrated fairness in terms of fairness concepts. Meanwhile, a few researchers have recently explored other concepts of fairness, such as Rawlsian maximin fairness. 

Although many efforts about fairness definitions have been made, there are still some issues. Firstly, the relationships among these fairness definitions, especially in recommender systems, lack adequate exploration. If these fairness definitions conflict, which definition is more important is also a problem. Consensus on what kind of fairness should be achieved in recommender systems is necessary. Note that people may have different fairness needs \cite{202115}, and consensus may not be the same in different scenarios. 
Besides, most studies concentrate on a single concept and target of fairness. Only a few recent studies \cite{202118,202009} attempt to achieve multiple fairness definitions simultaneously. If it is necessary to satisfy multiple fairness definitions, ensuring different fairness at the same time is also a question worth exploring.

\begin{table}[]
\caption{A lookup table for the reviewed fairness definitions in recommendation.}
\label{tab:def}
\centering
\resizebox{\textwidth}{!}{%
\begin{tabular}{c|l|l|l|l}
\toprule \toprule
\multicolumn{2}{l|}{\textbf{Fairness Definitions}}                                                                    & \textbf{Abbr.} & \textbf{Description} &  \textbf{Related Work}                                                                                                                                                                                                                                                    \\ \hline
\multicolumn{2}{l|}{Process Fairness}    & PF &  the allocation process should be fair                                                                 & \cite{202109,201904}                                                                                                                                                                                                                          \\ \hline \hline
\multicolumn{2}{l|}{Outcome Fairness}    & OF &  the allocation outcome should be fair                                                                  &   all the below                                                                                                                                                                                                                     \\ \hline
\multirow{2}{*}{\bigcell{c}{Grouped by \\ Target}}       & Individual Fairness       & IF & fairness should be guaranteed at the individual level & \bigcell{l}{\cite{202114,202115,202117,202118,202003,202004,202009,202010,202018,201907,202202} \\ \cite{202120,202006,202012,201916,201703,201704,201808,201912,201913,201806,202206,202122,202203}}                                                                                                                             \\ \cline{2-5} 
                                   & Group Fairness    & GF  & fairness should be guaranteed at the group level     & \bigcell{l}{\cite{202101,202102,202105,202107,202110,202112,202118,202003,202005,202008,202011,202205} \\ \cite{201907,201908,201801,201805,201809,201810,201811,201813,201814,201815,202204}\\ \cite{201707,201708,202103,202106,202108,202114,202116,202119,202013,202006} \\ \cite{201905,201906,202121,201702,201705,201818,201802,201820,202201,202017} \\ \cite{202014,202015,201816,201817,201819,202016,201902,201901,201909,201910}
                                  }\\ \cline{1-5} 
                                  \multirow{6}{*}{\bigcell{c}{Grouped by \\ Concept}} & Consistent Fairness       & CO & similar individuals / different groups should receive similar outcomes & \bigcell{l}{\cite{202101,202102,202103,202105,202106,202107,202108,202112,202114,202119} \\ \cite{202008,202015,202010,202012,202003,202017,202016,202018,201906,201907} \\ \cite{201801,201805,201806,201820,201810,201811,201809,201813,201814,201702,202203} \\ \cite{202013,202006,201912,201913,201708,202121,201910,202120,202004,202005,201908}\\
                                  \cite{201909,201704,201707}}                                                 \\ \cline{2-5} 
                                  &  Calibrated Fairness       & CA  & outcomes should be proportional to merits & \bigcell{l}{\cite{202110,202116,202006,202008,202011,202014,201901,201902,201905,201916, 202205} \\ \cite{201818,201819,201705,202201,201808,201815,201816,201817,202204}}                                                                                                                                      \\ \cline{2-5} 
                                  &   Counterfactual Fairness   & CF & \bigcell{l}{ individuals should have the same allocation outcome in the real world \\ as they do in the counterfactual world} & \cite{202115}                                                                                                                                                                                                                                        \\ \cline{2-5} 
                                  &  Rawlsian Maximin Fairness & RMF & the outcomes of the worst should be maximized & \cite{202117,202118,202122}                                                                                                                                                                                                                                 \\ \cline{2-5} 
                                  &  Envy-free Fairness         & EF & individuals should be free of envy & \cite{202009, 202202,201703,202206}                                                                                                                                                                                                                                 \\ \cline{2-5} 
                                  &   Maximin-Shared Fairness    & MSF & individuals / groups should get better outcomes than their maximin share & \cite{202009}                                                                                                                                                                                                                                        \\
\bottomrule \bottomrule
\end{tabular}%
}
\end{table}

\subsection{Relationships between Fairness and Other Concepts}
In this subsection, we discuss the relationship between fairness and some related concepts in recommender systems.

\textit{\textbf{Bias.}}
Bias is ubiquitous in recommender systems, which can exist in data, models, and outcomes \cite{200027}.
Bias may increase both outcome unfairness and process unfairness. For example, Zhu et al. \cite{202104} demonstrate theoretically that matrix factorization models suffer from popularity bias in the learning process, which causes popular items to be preferred when the true preferences are the same. Besides, the inductive bias of representation learning may tend to learn some sensitive information to increase the information contained in the representation, which may increase process unfairness. Thus, removing fairness-related biases in data and models is helpful in alleviating unfairness \cite{202107}. Besides, there are also some biases that are not related to fairness, such as position bias. In general, bias is more concerned with statistical issues, while fairness all reflects normative ideas about how a recommender system should be.

\textit{\textbf{Diversity.}}
Diversity in recommendation means the diversity of items in the recommendation list, which is closely related to user satisfaction \cite{200045}. For item fairness, improvements in item fairness are possible to increase diversity. It is because when optimizing item fairness, the recommendation list tends to contain more cold items as well as items from more categories \cite{202004,201805}, which means higher recommendation diversity. However, increasing diversity does not necessarily improve item fairness. The recommender system may recommend more popular items in each category, and cold items will still be treated unfairly. For user fairness, some studies find that existing methods to optimize recommendation diversity may exacerbate user unfairness \cite{201806}. Generally speaking, fairness is an evaluation criterion beyond diversity. Except for the fairness of accuracy, we can also consider the fairness of diversity \cite{202119}.

\textit{\textbf{Privacy.}}
Privacy requires that external attackers cannot obtain sensitive information about users through recommendation results or parameters of the recommendation model \cite{200046}. Compared with privacy, fairness is an internal perspective of the recommender system, with no consideration of external attackers. Nevertheless, some fairness definitions may imply privacy, such as process fairness and counterfactual fairness. Process fairness requires that the recommendation process should be as fair as possible, such as using fair representations. If we consider that fair representations should be independent of fairness-related attributes, then a fair representation will also satisfy privacy for these attributes. Moreover, in the counterfactual perspective, Li et al. \cite{202115} demonstrate that counterfactual fairness of users can be guaranteed by making user representations independent of fairness-related attributes. This implies that user representations satisfying privacy can guarantee counterfactual fairness.

\section{Views of Fairness in Recommendation}
The definitions of fairness introduced in Section 2 can be applied to any allocation process and are not limited to the recommendation. Whereas, in recommender systems, there exist multiple allocation processes corresponding to different fairness issues. In this section, to deepen the understanding of fairness, we present several views to classify fairness issues in the recommendation. These views and corresponding work are summarized in Table \ref{tab:views}.

{
\begin{table}[h]
\small
\caption{A lookup table for the reviewed fairness work from several views.}
\label{tab:views}
\centering
\begin{tabular}{l|l|l}
\toprule \toprule
\multicolumn{2}{l|}{\textbf{Fairness Views}}                       &       \textbf{Related Work}                                                                                                                                                                                                                                                                           \\ \hline
\multirow{3}{*}{Divided by Subject} & User            & \bigcell{l}{\cite{202105,202106,202107,202109,202112,202115,202119,202013,202010,202003,202018,201901,201904,201907,201909,202202} \\ \cite{201910,201913,201801,201802,201806,201811,201809,201813,201814,201702,201703,201704,202201,202206}}                                                                                                                 \\ \cline{2-3} 
                                    & Item            & \bigcell{l}{\cite{202101,202102,202103,202108,202110,202116,202117,202118,202120,202004,202005,202006,202015,202011,202012} \\ \cite{202014,202017,202016,201901,201902,201904,201905,201906,201907,201908,201912,201916,201815,201816,201817,202205,202204} \\ \cite{201805,201820,201808,201810,201811,201809,201814,201818,201819,201705,201707,201708,202121,202201,202203}}                                           \\ \cline{2-3} 
                                    & Joint           & \cite{202114,202008,202009,202122}                                                                                                                                                                                                                                                      \\ \hline
\multirow{2}{*}{Divided by Granularity}   & Single          & \bigcell{l}{\cite{202101,202103,202108,202115,202118,202005,202006,202015,202010,202016,202018,201902,201905,201906,201908,202205} \\ \cite{201913,201916,201816,201805,201810,201818,201819,201703,201704,201707,201708,202202,202206}}                                                                                                                                     \\ \cline{2-3} 
                                    & Amortized       & \bigcell{l}{\cite{202102,202105,202106,202107,202110,202112,202114,202116,202117,202119,202120,202004,202013,202008,202009}\\ \cite{202011,202012,202003,202014,202017,201901,201907,201909,201910,201912,201801,201815,201817,201806,201820} \\ \cite{201808,201811,201809,201813,201814,201702,201705,202121,202201,202204,202122,202203} }                                                       \\ \hline
\multirow{2}{*}{Divided by Optimization Object} & Treatment & \bigcell{l}{\cite{202101,202102,202103,202106,202108,202110,202114,202115,202116,202117,202118,202004,202005,202013,202006} \\ \cite{202008,202009,202015,202010,202011,202012,202003,202014,202017,202018,201901,201902,201905,201908,201909} \\ \cite{201910,201912,201913,201916,201815,201816,201817,201805,201806,201820,201808,201810,201811,201809,201818} \\
\cite{201819,201703,201704,201705,201707,201708,202121,202205,202206,202204,202122,202203}} \\ \cline{2-3} 
                                    & Impact    & \bigcell{l}{\cite{202101,202105,202107,202112,202116,202119,202008,202014,202017,202016,201901,201906,201907,201801,201813,201814} \\ \cite{201702,202201,202202}}                                                                                                                                                                  \\
\bottomrule \bottomrule
\end{tabular}%
\end{table}
}

\subsection{Subject}
We refer to the subjects (e.g., individual $i$ and group $G_i$ in Section 2) receiving allocation in the allocation process as \textit{fairness subject}, which corresponds to "Fair for Who."  As there are different kinds of subjects in recommendation, fairness can be divided into \textbf{item fairness}, \textbf{user fairness}, and \textbf{joint fairness}. As demonstrated in Table \ref{tab:views}, previous work mainly concentrates on item fairness and user fairness, and only a little work aims to improve joint fairness.

Item fairness concerns whether the recommendation treats items fairly, such as similar prediction errors for ratings of different types of items \cite{201907} or allocating exposure to each item proportional to its relevance \cite{201808}. If the recommendation treats different items unfairly, the providers of these discriminated items may lack positive feedback and leave the platform. Calibrated fairness is frequently applied to item fairness, while there is little work about calibrated fairness of users, probably because items are easily associated with concepts such as value and quality. The value of an item is often measured by its relevance to users \cite{202014} or the number of interactions in history \cite{202102}. Note that some researchers \cite{201701,201903} divide the subjects into consumer fairness and provider fairness. In contrast, we divide the subjects into user fairness and item fairness here, as provider fairness can be considered a kind of item fairness at the group level, where groups are divided according to providers.

User fairness concerns whether the recommendation is fair to different users, such as similar accuracy for different groups of users \cite{201813} or similar recommendation explainability across different users \cite{202003}. If the recommendation cannot be fair to users, it may lose users with specific interests. The most commonly used fairness definition in user fairness is consistent fairness, as it is often believed that different people are similar and should not be treated differently. However, here are some particular scenarios where fairness means treating people differently. For example, premium members should get better recommendations than standard members \cite{201901}. Moreover, there are some differences between user fairness in group recommendations and general recommendations. User fairness in general recommendations concerns all users \cite{202112}, while group recommendations only care about the users in the group receiving the recommendation \cite{201703}.

Joint fairness concerns whether both users and items are treated fairly \cite{202114}. In most recommendation scenarios, it is necessary to consider joint fairness, as user fairness and item fairness are vital to most recommender systems. It is worth noting that user fairness and item fairness can conflict with each other. When item fairness is improved, user fairness must worsen or remain the same \cite{202114}, making joint fairness a challenging problem.

In addition to users and items, a few other stakeholders may exist in recommender systems. Their fairness issues have recently received attention from some researchers \cite{201903}.

\subsection{Granularity}
We refer to the granularity of the allocation process as \textit{fairness granularity}. Fairness in recommendation can be further divided into \textbf{single fairness} and \textbf{amortized fairness}.

A single recommendation list can be considered as the minimum allocation process in the recommendation, which corresponds to \textbf{single fairness}. Single fairness requires that the recommender system meets fairness requirements each time it generates a single recommendation list. In other words, the outcomes $C_h(\cdot)$ are only related to a single recommendation, and each recommendation should satisfy the specific fairness definition. For example, for item fairness, different types of items in a single recommendation list should satisfy the fair distribution \cite{201816}. For user fairness, a single recommendation list should be similarly relevant for different users in the group recommendation \cite{201703}. 

However, requiring every single recommendations list to be fair may be difficult and performance-damaging. An alternative is that we require the recommendations to be fair on the cumulative level, which is called \textbf{amortized fairness} \cite{201808}. Amortized fairness requires that the cumulative effect of multiple recommendation lists is fair, while a single recommendation list in them may be unfair. In other words, the outcomes $C_h(\cdot)$ are related to multiple recommendations.

For example, suppose we expect the exposure of books by male authors and books by female authors to be close in book recommendations. Single fairness requires that each recommendation list has approximately the same number of books by male authors as by female authors. In contrast, amortized fairness will only require that the system recommends approximately the same number of books by male authors as by female authors in all recommendations over time (e.g., within a day).

As demonstrated in Table \ref{tab:views}, previous studies concentrate on amortized fairness, which is probably because single fairness is not achievable in some scenarios \cite{201808}. Existing work \cite{202014,202110} often uses the \textit{average} value as the cumulative effect, such as the average exposure of a group across multiple recommendation lists. However, even if the \textit{average} values are the same, the \textit{variance} may be different, which may also be unfair. A high variance may mean that the recommendation performance is not stable and may bring more negative experiences to users. Nevertheless, no previous work has been done that takes \textit{variance} into consideration.

\subsection{Optimization Object}
We refer to the aspect in which we are concerned about the allocation for subjects as \textit{optimization object}, which is consistent with how the value function $V(\cdot)$ is defined in Section 2. There are many kinds of optimization objects, containing exposure and hit ratio of items \cite{202014}, and accuracy of recommendations for users \cite{202119}. According to whether to consider the impact of allocation \cite{200007,200008}, they can be divided into two main types, i.e., \textbf{treatment-based fairness} and \textbf{impact-based fairness}. Treatment-based fairness only considers whether the treatments of the recommender system are fair or not, such as the predicted scores to different users \cite{201811} and the allocated exposure to different items \cite{202009}.  
In contrast, impact-based fairness takes the impact caused by recommendations (i.e., user feedback) into account. Taking item fairness as an example, in the Top-N ranking task, treatment-based fairness may require that the exposure of different items conforms to a fair distribution \cite{201905}. In contrast, impact-based fairness may require that the CTR of different items conforms to a fair distribution \cite{202014}.

As shown in Table \ref{tab:views}, most previous studies have focused on treatment-based fairness. It may be because it is more difficult to consider impact-based fairness as we cannot control user feedback directly. While most work only focuses on treatment-based fairness or impact-based fairness, it is also necessary to consider both impact-based fairness and treatment-based fairness together. Using item fairness as an example, on the one hand, if we only consider exposure without concerning the accuracy of recommendations, then there is a risk that the recommender system tends to recommend discriminated items to some inactive users. Although the exposure increases, the drop in click-through rate may instead lead to a loss of confidence of the provider. On the other hand, if we only consider the accuracy without considering the exposure of the recommendation, it may lead the recommender system to reduce the exposure chance of the discriminated items to reduce the decrease of the accuracy, which is also unfavorable for the discriminated items. Therefore, it is necessary to consider both impact-based fairness and treatment-based fairness.

\section{Measurements of Unfairness in Recommendation}
\subsection{Overview of Fairness Metrics}
We introduce some widely used metrics for fairness in the recommendation, as shown in Table \ref{tab:measurements}. Since there are different fairness definitions, the measurements of unfairness are not the same. Moreover, as the characteristics of fairness issues mentioned in Section 3 also affect the design and choice of fairness metrics, different metrics have different scopes of application, which are also marked in Table \ref{tab:measurements}.

As demonstrated in Table \ref{tab:measurements}, most fairness metrics are proposed for outcome fairness as it is the focus of most work, where more metrics for consistent fairness and calibrated fairness. Thus, we mainly present the corresponding metrics for these two fairness definitions in sections 4.2 and 4.3, respectively, and show all the others in section 4.4.

When selecting fairness metrics based on definitions, it is important to note that different metrics do not have the same scope of application. For consistent fairness, \textit{Absolute Difference}, \textit{Variance}, and \textit{Gini coefficient} are commonly used measurements at the two-group, multi-group, and individual levels. These three metrics have a wide range of applicability to different subjects, granularity, and optimization objects. For calibrated fairness, \textit{KL-divergence} and \textit{L1-norm} are common measurements for multi-group and individual fairness. These two metrics also have broad applicability. Due to many groups in the group-level calibrated fairness studies, there are no metrics specifically designed for the two group situations. These common metrics are generic and can be used for both users and items but are relatively coarse-grained. They have two main drawbacks: 

(1) These common metrics typically use the first-order moment like the average to describe groups, ignoring higher-order information; 

(2) These metrics do not consider the characteristics of user fairness and item fairness. 

In order to address the first point, some researchers \cite{202008,201811} use statistical tests like \textit{KS statistic} or \textit{ANOVA} that consider the population distribution. For the second point, for users, some researchers \cite{201702} consider user fairness on each item and then aggregate them. For items, some researchers \cite{201708,201905} consider unfairness across different positions and then aggregate them. Although limited in application, these metrics could be more proper for specific fairness issues. Specific details of these metrics are described below.

Since the metrics for different fairness definitions are not the same, we next present the corresponding metrics based on the fairness definitions. The meanings of the commonly used symbols are shown in Table \ref{tab:notations}.

\begin{table}[h]
\caption{A lookup table for the reviewed fairness measurements with the order of Def. and the Target. "$\checkmark$" denotes the presence of existing work using the metric under the corresponding conditions. "-" means that there is no work to use the metric in the corresponding condition, but the metric could theoretically be used in the corresponding condition as well. "$\times$" indicates that the metric is not theoretically available for the corresponding condition. The abbreviations of the definitions are shown in Table \ref{tab:def}. We use "1" to denote those measurements without a name in the original paper and "2" to denote those measurements with the original name. }
\label{tab:measurements}
\centering
\resizebox{\textwidth}{!}{%
\begin{tabular}{l|l|p{1cm}<{\centering}p{1cm}<{\centering}c|cc|cc|cc|l}
\toprule \toprule
\multirow{2}{*}{\textbf{Metric Name}}   & \multirow{2}{*}{\textbf{Def.}}                                                     & \multicolumn{3}{c|}{\textbf{Target}}                                            & \multicolumn{2}{c|}{\textbf{Subject}} &
\multicolumn{2}{c|}{\textbf{Granularity}} &  \multicolumn{2}{c|}{\textbf{Optim. Object}}                & \multirow{2}{*}{\textbf{Related Work}} \\[0.1em]
\cline{3-11} \rule{0em}{2.5ex}
    &   & Two groups & More groups & \multirow{2}{*}{Ind.} & \multirow{2}{*}{User}    & \multirow{2}{*}{Item}   & \multirow{2}{*}{Single} & \multirow{2}{*}{Amortized}  & \multirow{2}{*}{Treat.} & \multirow{2}{*}{Impact} &                               \\ 
\midrule \midrule
Absolute Difference\footnote{We use "1" to denote those measurements without name in the original paper. }            & CO                                                                       & \checkmark             & $\times$         & $\times$        & \checkmark     & \checkmark & \checkmark &  \checkmark  & \checkmark             & \checkmark          &   \cite{201811,202003,202101,202112}                            \\ \hline
KS statistic\footnote{We use "2" to denote those measurements with the original name. }                   & CO                                                                       & \checkmark             & $\times$         & $\times$        & \checkmark     & -   & -  & \checkmark    & \checkmark             & -            &        \cite{201811,201814}                       \\ \hline
rND$^2$                            & CO                                                                       & \checkmark             & $\times$         & $\times$        & $\times$     & \checkmark  &  \checkmark  &   $\times$     & \checkmark             & $\times$          &  \cite{201708}                             \\ \hline
rKL$^2$                            & CO                                                                       & \checkmark             & $\times$         & $\times$        & $\times$     & \checkmark &  \checkmark  &   $\times$    & \checkmark             & $\times$          &  \cite{201708}                             \\ \hline
rRD$^2$                            & CO                                                                       & \checkmark             & $\times$         & $\times$        & $\times$     & \checkmark &  \checkmark  &   $\times$    & \checkmark             & $\times$          &  \cite{201708}                             \\ \hline
Pairwise Ranking Accuracy Gap$^2$  & CO                                                                       & \checkmark             & $\times$         & $\times$        & $\times$     & \checkmark      &  \checkmark    &   $\times$     & $\times$             & \checkmark          &  \cite{201906,202101}                             \\ \hline
Value Unfairness$^2$               & CO                                                                       & \checkmark             & $\times$         & $\times$        & \checkmark     & $\times$   &  $\times$  &    \checkmark       & $\times$             & \checkmark          &      \cite{201702,201801}                         \\ \hline
Absolute Unfairness$^2$            & CO                                                                       & \checkmark             & $\times$         & $\times$        & \checkmark     & $\times$ &  $\times$  &    \checkmark    & $\times$             & \checkmark          &    \cite{202107,201801,201702}                           \\ \hline
Underestimation Unfairness$^2$     & CO                                                                       & \checkmark             & $\times$         & $\times$        & \checkmark     & $\times$  &  $\times$  &    \checkmark    & $\times$             & \checkmark          &     \cite{201702,201801}                          \\ \hline
Overestimation Unfairness$^2$      & CO                                                                       & \checkmark             & $\times$         & $\times$        & \checkmark     & $\times$ &  $\times$  &    \checkmark       & $\times$             & \checkmark          &      \cite{201702,201801}                         \\ \hline
Variance$^2$                       & CO                                                                       & -               & \checkmark         & \checkmark        & \checkmark     & \checkmark    & \checkmark &  \checkmark   & \checkmark             & \checkmark          &          \cite{201704,201907,202114}                     \\ \hline
Min-Max Difference$^1$             & CO                                                                       & -               & \checkmark         & \checkmark        & \checkmark     & \checkmark  &  \checkmark   &  \checkmark      & \checkmark             & -            &       \cite{202103,202018}                        \\ \hline
F-statistic of ANOVA$^2$           & CO                                                                       & -               & \checkmark         & $\times$        & \checkmark     & \checkmark    & -  &  \checkmark   & -               & \checkmark          &          \cite{202008}                     \\ \hline
Gini coefficient$^2$               & CO                                                                       & -               & -           & \checkmark        & \checkmark     & \checkmark     & -    &   \checkmark       & \checkmark             & -            &             \bigcell{l}{\cite{201806,202004,202003,202102} \\ \cite{202121,202122,202203}}                  \\ \hline
Jain's index$^2$                   & CO                                                                       & -               & -           & \checkmark        & \checkmark     & \checkmark & \checkmark  &  \checkmark    & \checkmark             & -            &             \cite{201704,202012}                  \\ \hline
Entropy$^2$                        & CO                                                                       & -               & -           & \checkmark        & -       & \checkmark    &  - & \checkmark  & \checkmark             & -            &                   \cite{202004,202009,202121}            \\ \hline
Min-Max Ratio$^2$                  & CO                                                                       & -               & -           & \checkmark        & \checkmark     & - &  \checkmark  &  -   & \checkmark             & -            &                   \cite{201704,202010}            \\ \hline
Least Misery$^2$                   & CO \& RMF & -               & -           & \checkmark        & \checkmark     & -  &  \checkmark  & -  & \checkmark             & -            &                  \cite{202010,201913,201704}             \\ \hline
MinSkew$^2$                        & CA                                                                       & -               & \checkmark         & $\times$        & -       & \checkmark   &  \checkmark   &  -   & \checkmark             & -            &        \cite{201905}                        \\ \hline
MaxSkew$^2$                        & CA                                                                       & -               & \checkmark         & $\times$        & -       & \checkmark  &  \checkmark   &  -    & \checkmark             & -            &        \cite{201905}                      \\ \hline
KL-divergence$^2$                  & CA                                                                       & -               & \checkmark         & -          & -       & \checkmark    &  \checkmark   &  \checkmark   & \checkmark             & -            &           \cite{201816,202008,202201,202204}                 \\ \hline
NDKL$^2$                           & CA                                                                       & -               & \checkmark         & -          & $\times$     & \checkmark   &  \checkmark    &  $\times$   & \checkmark             & -            &        \cite{201905}                       \\ \hline
JS-divergence$^2$                  & CA                                                                       & -               & \checkmark         & -          & -       & \checkmark    &  -   &    \checkmark    & \checkmark             & -            &           \cite{201705}                    \\ \hline
Overall Disparity$^1$              & CA                                                                       & -               & \checkmark         & -          & -       & \checkmark   &  -  &  \checkmark   & \checkmark             & \checkmark          &         \cite{202014,202110}                      \\ \hline
Generalized Cross Entropy$^2$      & CA                                                                       & \checkmark             & \checkmark         & -          & \checkmark     & \checkmark  &  -   &  \checkmark    & \checkmark             & \checkmark          &           \cite{201901,202201}                    \\ \hline
L1-norm$^2$                        & CA                                                                       & -               & \checkmark         & \checkmark        & -       & \checkmark     &   \checkmark  &  \checkmark  & \checkmark             & -            &           \cite{201808, 201912,202108}                    \\ \hline
Proportion of Envy-free Users$^1$   & EF                                                                         & $\times$             & $\times$         & \checkmark        & \checkmark     & $\times$      &  \checkmark   &  $\times$  & \checkmark             & -            &             \cite{201703,202206}                  \\ \hline
Mean Average Envy$^1$              & EF                                                                         & $\times$             & $\times$         & \checkmark        & \checkmark     & $\times$       &  $\times$    & \checkmark  & \checkmark             & -            &             \cite{202009}                  \\ \hline
Classification-based Metrics$^1$   & CF \& PR      & $\times$            & $\times$         & \checkmark        & \checkmark     & \checkmark   &       $\times$   & $\times$  & $\times$            & $\times$          &         \cite{202115,202109,201904}                      \\ \hline
Bottom N Average$^1$               & RMF                                                                 & $\times$            & -           & \checkmark        & -       & \checkmark  & -       & \checkmark   & -               & \checkmark          &           \cite{202117,202122}                    \\ 
\hline
Fraction of Satisfied Producer$^1$ & MSF                                                                   & $\times$             & $\times$         & \checkmark        & $\times$     & \checkmark  & -            & \checkmark   & \checkmark             & $\times$          &          \cite{202009}                     \\
\bottomrule \bottomrule
\end{tabular}%
}
\end{table}

\begin{table}[]
\caption{Notations and Explanations of Common Variables}
\label{tab:notations}
\begin{tabular}{l|l}
\toprule \toprule
\textbf{Notation} & \textbf{Explanation}              \\
\hline
$n$     & number of users                        \\
$m$      & number of items                        \\
$k$     & length of recommendation list                                         \\
$\hat{r}_{u,i}$       & prediction for user $u$ and item $i$                       \\
${r}_{u,i}$          &  feedback of user $u$ to item $i$      \\
$\mathcal{U} = \{ u_1,...,u_n \} $    & the whole set of users                        \\
$\mathcal{I} = \{ i_1,...,i_m \} $   & the whole set of items                        \\ 
$\mathcal{L} = \{l_{u_1},...,l_{u_n}\}$    & the whole set of recommendation lists, $|l_{u_i}|= k$ \\
$\mathcal{R} = \{ r_{u,i} \}$   & the whole set of feedback\\ 
$\mathcal{V}$ & the whole set of individuals or groups, which can be either users or items \\
$f(\cdot)$ & the utility function for individuals or groups \\
\bottomrule \bottomrule
\end{tabular}
\end{table}

\subsection{Metrics for Consistent Fairness (CO)}

As mentioned in Section 2, current work on consistent fairness in recommendation requires that all individuals or groups should be treated similarly. Therefore, the corresponding measurements mainly measure the inconsistency of the utility distribution. Most metrics apply to both user fairness and item fairness. They consider the utility of each individual or group as a number and then measure the inconsistency of these numbers. Due to many metrics on consistent fairness and that early studies concentrate on situations where only two groups exist, we will present these metrics in the order of metrics for two groups, multiple groups, and individuals.

\textit{\textbf{Absolute Difference.}} Absolute Difference (AD) is the absolute difference of the utility between the protected group $G_0$ and the unprotected group $G_1$. For user, the group utility $f(G)$ is often defined as the average predicted rating \cite{201811} or the average recommendation performance in the group $G$ \cite{202112,202003}. For item, the group utility $f(G)$ can be defined as the whole exposure in the recommendation lists for the group $G$ \cite{202101}. The lower the value, the fairer the recommendations.
\begin{equation} 
AD = | f(G_0) - f(G_1) |
\end{equation}

\textit{\textbf{KS statistic.}}
Kolmogorov-Smirnov statistic is a nonparametric test used to determine the equality of two distributions. It measures the area difference between two empirical cumulative distributions of the utilities for groups. The utilities are often defined as the predicted ratings in the group \cite{201811,201814}. Compared to $AD$ using the average utility, KS statistic can measure the high-order inconsistency. The lower the value, the fairer the recommendations.
\begin{equation} 
KS = |\sum_{i = 1}^T l \times \frac{\mathcal{G}(R_0,i)}{|R_0|} - \sum_{i = 1}^T l \times \frac{\mathcal{G}(R_1,i)}{|R_1|}| 
\end{equation}
Here $T$ is the number of intervals in the empirical cumulative distribution, $l$ is the size of each interval, $\mathcal{G}(R_0,i)$ is the number of utilities of the group $G_0$ that are inside the $i$-$th$ interval.

\textit{\textbf{rND, rKL and rRD.}}
rND, rKL and rRD measure item exposure fairness for a ranking $\tau$ \cite{201708}. Unlike previous metrics, these metrics take the exposure position into account, calculating the normalized discounted cumulative unfairness similar to NDCG. Experiments show that rKD is smoother and more robust than rRD, and that rRD has limited application scope. The lower the value, the fairer the recommendations are for these metrics. 
\begin{equation} 
rND = \frac{1}{Z}\sum_{i = 10,20,...}^N \frac{1}{\log_2 i} |\frac{|S^+_{1...i}|}{i} - \frac{|S^+|}{N}|
\end{equation}
\begin{equation} 
rKL = \frac{1}{Z}\sum_{i = 10,20,...}^N \frac{1}{\log_2 i} (\frac{|S^+_{1...i}|}{i}\log\frac{\frac{|S^+_{1...i}|}{i}}{\frac{|S^+|}{N}} + \frac{|S^-_{1...i}|}{i}\log\frac{\frac{|S^-_{1...i}|}{i}}{\frac{|S^-|}{N}})
\end{equation}
\begin{equation} 
rRD = \frac{1}{Z}\sum_{i = 10,20,...}^N \frac{1}{\log_2 i} |\frac{|S^+_{1...i}|}{|S^-_{1...i}|} - \frac{|S^+|}{|S^-|}|
\end{equation}
Here the normalizer Z is the highest possible value of corresponding measurements, $|S^+_{1...i}|$ is the number of the protected group in the top-i of the ranking $\tau$, $S^+$ is the number of the unprotected group in the whole ranking.

\textit{\textbf{Pairwise Ranking Accuracy Gap.}}
Pairwise Ranking Accuracy Gap (PRAG) measures item unfairness in the pairwise manner \cite{202101,201906}. Unlike previous metrics focusing on exposure or click-through rate, PRAG measures the unfairness of pairwise ranking accuracy, and it is calculated on data from randomized experiments. The lower the value, the fairer the recommendations.
\begin{equation} 
PRAG = |PairAcc(I_1 > I_2 | q) - PairAcc(I_1 < I_2 | q)|
\end{equation}
\begin{equation} 
PairAcc(I_1 > I_2 | q) = P(f(x_i) > f(x_j) | y_i > y_j , i \in I_1, j \in I_2)
\end{equation}
Here $PairAcc$ represents the ranking accuracy for a pair of items $x_i, x_j$ from different groups $I_1,I_2$. $f(x_i)$ and $f(x_j)$ are the predicted score for the recommendation query $q$. $y_i$ and $y_j$ are the true feedback, which are collected through randomized experiments.  

\textit{\textbf{Value Unfairness and its variants.}}
Value unfairness is proposed to measure inconsistency in signed prediction error between two user groups \cite{201702}. There are three variants of Value unfairness. Absolute Unfairness measures the inconsistency of absolute prediction error, while Underestimation Unfairness and Overestimation Unfairness measure inconsistency in how much the predictions underestimate and overestimate the true ratings, respectively. The lower the value, the fairer the recommendations.
\begin{equation} 
U_{val} = \frac{1}{m} \sum_{i = 1}^m | (E_0[\hat{r}]_i - E_0[r]_i) - (E_1[\hat{r}]_i - E_1[r]_i) | 
\end{equation}
\begin{equation}
U_{abs} = \frac{1}{m} \sum_{i = 1}^m | |E_0[\hat{r}]_i - E_0[r]_i| - |E_1[\hat{r}]_i - E_1[r]_i| |
\end{equation}
\begin{equation} 
U_{under} =  \frac{1}{m} \sum_{i = 1}^m | max(0,E_0[r]_i - E_0[\hat{r}]_i) - max(0,E_1[r]_i - E_1[\hat{r}]_i) | 
\end{equation}
\begin{equation}
U_{over} = \frac{1}{m} \sum_{i = 1}^m | max(0,E_0[\hat{r}]_i - E_0[r]_i) - max(0,E_1[\hat{r}]_i - E_1[r]_i)| 
\end{equation}
Here $E_0[\hat{r}]_i$ is the average predicted score for the $i$-$th$ item from group 0, and $E_0[r]_i$ is the average rating for the $i$-$th$ item from group 0.
\newline

The above metrics are only applicable to measure inconsistency between two groups. In the following, we present the metrics to measure unfairness for \textbf{three or more groups}. It is worth noting that since we can consider individual fairness as a special case of group fairness (i.e., each individual belongs to a unique group), theoretically, these group fairness metrics below can also apply to individual fairness. However, in practice, the common metrics for individual and group fairness are different.

\textit{\textbf{Variance.}} Variance is a commonly used metrics for dispersion, which is applied to both group-level \cite{202114,201907} and individual-level \cite{202114,201907,201704}. The utility can be the rating prediction error \cite{201907}, the predicted recommendation satisfaction for a single user \cite{201704,202114} and the average exposure for an item group \cite{202114}. The lower the value, the fairer the recommendations.
\begin{equation}
Variance = \frac{1}{|\mathcal{V}|^2}\sum_{v_x \neq v_y} (f(v_x) - f(v_y))^2  
\end{equation}

\textit{\textbf{Min-Max Difference.}} Min-Max Difference (MMD) is the difference between the maximum and the minimum of all allocated utilities. This metric is used to measure the inconsistency of the average exposure for multiple item groups \cite{202103}, and the disagreement for users in group recommendation at the individual level \cite{202018}. The lower the value, the fairer the recommendations.
\begin{equation} 
MMD = max\{f(v),\forall v \in \mathcal{V}\} - min\{f(v),\forall v \in \mathcal{V}\}
\end{equation}

\textit{\textbf{F-statistic of ANOVA.}} 
The one-way analysis of variance (ANOVA) is used to determine any statistically significant differences between the mean values of three or more independent groups. Its F-statistic can be considered a fairness measurement. The utility can be the rating prediction error for a single rating \cite{202008}. The lower the value, the fairer the recommendations.
\begin{equation} 
F = \frac{MST}{MSE}
\end{equation}
\begin{equation}
MST = \frac{\sum_i |v_i| \times (\overline{v}_i - \overline{v})^2}{|\mathcal{V}| - 1}
\end{equation}
\begin{equation}
MSE = \frac{\sum_i \sum_{j \in v_i} (f(ind_j) - \overline{v}_i)^2}{\sum_{v \in \mathcal{V}}|v| - |\mathcal{V}|}
\end{equation}
Here $f(ind_j)$ is the utility of an individual belong to $v_i$, $\overline{v}_i$ is the mean utility of group $v_i$, $\overline{v}$ is the mean utility of all individuals.
\newline

In the following, we present some metrics commonly used for \textbf{individual fairness}. Note that in addition to the metrics below, $Variance$ above is also often used to measure individual fairness.

\textit{\textbf{Gini coefficient.}} Gini coefficient is widely used in sociology and economics to measure the degree of social unfairness \cite{201806,202004,202003,202102,202121}. To our knowledge, it is also the most commonly used metric for consistent individual fairness. The utility can be the predicted relevance for a user \cite{201806,202003} or the exposure for an item \cite{202004,202102,202121}. The lower the value, the fairer the recommendations.
\begin{equation}
Gini = \frac{\sum_{v_x,v_y \in \mathcal{V}}|f(v_x) - f(v_y)|}{2|\mathcal{V}| \sum_v f(v)}
\end{equation}

\textit{\textbf{Jain's index.}} Jain's index \cite{200052} is commonly used to measure unfairness in network engineering. Some studies use it to measure the inconsistency of predicted user satisfaction in group recommendations \cite{201704} and the inconsistency of item exposure \cite{202012}. The higher the value, the fairer the recommendations.
\begin{equation}
Jain = \frac{(\sum_v f(v))^2}{|\mathcal{V}|\cdot\sum_v f(v)^2} 
\end{equation}

\textit{\textbf{Entropy.}} Entropy is often used to measure the uncertainty of a system. In recommendation, it is used to measure the inconsistency of item exposure \cite{202004,202009,202121}. The lower the value, the fairer the recommendations.
\begin{equation}
Entropy = - \sum_{v \in \mathcal{V}} p(v) \cdot \log p(v) 
\end{equation}

\textit{\textbf{Min-Max Ratio.}} Min-Max Ratio is the ratio of the minimum to the maximum of all allocated utility. Some studies \cite{201704,202010} use it to measure the inconsistency of the predicted user satisfaction in group recommendation. The higher the value, the fairer the recommendations.
\begin{equation} 
MinMaxRatio = \frac{min\{f(v),\forall v \in \mathcal{V}\}}{max\{f(v),\forall v \in \mathcal{V}\}} 
\end{equation}

\textit{\textbf{Least Misery.}} Least Misery is the minimum of all allocated utility. It is also a commonly used fairness metric in group recommendation \cite{202010,201913,201704}. The higher the value, the fairer the recommendations.
\begin{equation} 
LeastMisery =  min\{f(v),\forall v \in \mathcal{V}\} 
\end{equation}

\subsection{Metrics for Calibrated Fairness (CA)}
Calibrated fairness requires defining the merit of an individual or group. We denote $Merit(\cdot)$ as a merit function that measures the merit of an individual or group.
We can calculate the fair distribution of the allocation based on $Merit(\cdot)$, i.e., the proportion of the individual's or group's allocation to the total allocation in the fair case, i.e., $p_f(v_i) = \frac{Merit(v_i)}{\sum_j Merit(v_j)}$. We can also calculate the proportion of the total allocation for an individual or group in the current situation, i.e., $p(v_i) = \frac{f(v_i)}{\sum_j f(v_j)}$. Most measurements of calibrated fairness measure the difference between the distribution of utilities $p$ and the distribution of merits $p_f$.

Since all the group fairness metrics in calibrated fairness can be applied to multiple groups, we will present them in the order of group fairness and individual fairness.

\textit{\textbf{MinSkew} and \textbf{MaxSkew}.} The deviation (Skew) on a certain group $v$ can be defined as $\log(\frac{p_f(v)}{p(v)})$. And then, we can define the min-skew and the max-skew as follows. Here the utility can be the exposure of the item group, while the $p_f$ is a predefined distribution \cite{201905}. For MinSkew, the higher the value, the fairer the recommendations. For MaxSkew, the lower the value, the fairer the recommendations.
\begin{equation}
Min-Skew = min\{\log(\frac{p_f(v)}{p(v)}), v \in \mathcal{V}\}
\end{equation}
\begin{equation}
Max-Skew = max\{\log(\frac{p_f(v)}{p(v)}), v \in \mathcal{V}\}
\end{equation}

\textit{\textbf{KL-divergence.}} KL-divergence measures how one probability distribution is different from the other. It can be used to measure the difference between $p_f$ and $p$. Here the utility can be the exposure of the item group, while the $p_f$ can be calculated by the group's historical exposure \cite{201816,202008,202201,202204}. The lower the value, the fairer the recommendations.
\begin{equation} 
D_{KL}(p,p_f) = \sum_{v \in \mathcal{V}} p(v)\log \space \frac{p(v)}{p_f(v)} 
\end{equation}

\textit{\textbf{NDKL.}}
NDKL is an item unfairness measure based on KL-divergence \cite{201905}. It computes the KL-divergence for each position and then obtains a normalized discounted cumulative value. The lower the value, the fairer the recommendations.
\begin{equation}
NDKL@K = \frac{1}{Z} \sum_i^K \frac{1}{\log(i+1)}D_{KL}^i 
\end{equation}
Here the normalizer Z is computed as the highest possible value, and $D_{KL}^i$ is the KL-divergence of the top-i ranking.

\textit{\textbf{JS.}} Like KL-divergence, JS-divergence also measures how one probability distribution differs from the other. Some work \cite{201705} uses JS-divergence as a metric instead of KL-divergence as it is symmetrical while KL-divergence is asymmetrical. The lower the value, the fairer the recommendations.
\begin{equation}
D_{JS}(p,p_f) = \frac{1}{2}(D_{KL}(p,\frac{1}{2}(p_f + p)) + D_{KL}(p_f,\frac{1}{2}(p_f + p)) )
\end{equation}

\textit{\textbf{Overall Disparity.}} Overall disparity measures the average disparity of the proportion of the utility and merit among different groups. The utility can be exposure-based or click-based \cite{202014,202110}. The lower the value, the fairer the recommendations.
\begin{equation}
OD = \frac{2}{|V|(|V| - 1)}\sum_{i = 0}^{|V|}\sum_{j = i + 1}^{|V|} || \frac{p(v_i)}{p_f(v_i)} - \frac{p(v_j)}{p_f(v_j)} ||
\end{equation}

\textit{\textbf{Generalized Cross Entropy.}} Generalized cross entropy \cite{201901, 202201} also measures how one probability distribution is different from the other. The higher the value, the fairer the recommendations.
\begin{equation} 
GCE = \frac{1}{\alpha (1- \alpha)} \left[ \sum_{v \in \mathcal{V}} p_f^\alpha(v)p^{(1 - \alpha)}(v) - 1 \right]
\end{equation}
Here $\alpha$ is a hyperparameter.
\newline

In the following, we present calibrated fairness measures frequently used at the individual level.

\textit{\textbf{L1-norm.}} L1-norm is the sum of the magnitudes of the vectors in a space. Some researchers \cite{201808, 201912,202108} treat the merit and utility distributions as vectors and then use the L1-norm to calculate the distance between the vectors. This metric is often used for individual-level measurement \cite{201808,201912}, and there is also work \cite{202108} that uses it to measure group-level unfairness. The lower the value, the fairer the recommendations.
\begin{equation}
L1-norm = \sum_{v \in \mathcal{V}} |p(v) - p_f(v)|
\end{equation}

It is worth noting that some measures of calibrated fairness and consistent fairness are interconvertible. Theoretically, for a calibrated fairness measurement, if we set $p_f$ to a uniform distribution, it can become a measurement for consistent fairness. On the other hand, for a consistent fairness measurement which contains $f(v)$, we can set $f(v)$ to $\frac{p(v)}{p_f(v)}$, then it become a calibrated fairness measurement.

\subsection{Metrics for Other Fairness Definitions}
\subsubsection{Metrics for Envy-free Fairness (EF)}
Envy-free fairness requires a definition of envy, which can be different in different scenarios. In group recommendations, different users in the group receive the same recommendations. Serbos \cite{201703} defines envy as follow:

\textit{\textbf{Envy-freeness (in group recommendation).}} Given a group $G$, a group recommendation package $P$, and a parameter $\delta$, we say that a user $u \in G$ is envy-free for an item $i \in P$ if $r_{u,i}$ is in the top-$\delta$\% of the preferences in the set $\{r_{v,i} : v \in G\}$.

This envy definition can be applied to a single item. This definition means that a user $u$ feels envy on an item if at least $\delta$\% users in the group like this item more than $u$. It is impossible for all users in a group to be envy-free (i.e., the user is envy-free for all items in the package). In practice, m-envy-free is often used, which means that the user in the group is envy-free for at least $m$ items.

A measurement for envy-free fairness can be the proportion of m-envy-free users:
\begin{equation} 
F = \frac{|G_{ef}|}{|G|} 
\end{equation}
where $|G_{ef}|$ is the number of m-envy-free users. The higher the value, the fairer the recommendations.
~\\

In general recommendations, different users receive different recommendations. Patro et al. \cite{202009} define envy-freeness as follow:

\textit{\textbf{Envy-freeness(in general recommendation).}} Given a utility metrics $f$ and all the recommendation lists $\mathcal{L}$, we say that a user $u$ is envy-free for a user $v$ if and only if $f(l_v,u) \ge f(l_u,u)$ and the degree of envy can be defined as $max(f(l_v,u) - f(l_u,u),0)$. Here $f(l,u)$ is the predicted relevance sum for the user $u$ with the recommendation list $l$. 

This envy definition is applied to each pair of users. Unlike envy in group recommendations, this definition does not involve the third user. Moreover, it is feasible to make all users envy-free with utility metrics properly chosen.

The average of envy among users can be a measurement of envy-free fairness:
\begin{equation}
Envy(\mathcal{U}) = \frac{1}{n\cdot(n-1)} \sum_{u_i, u_j, u_i \neq u_j}{envy(u_i,u_j)} 
\end{equation}
where $envy(u_i,u_j) = max(f(l_i,u_i) - f(l_j,u_i),0)$. The lower the value, the fairer the recommendations.

The two metrics above utilize predicted preferences that are often not consistent with users' true preferences. Envy-free fairness based on true preferences requires answering counterfactual questions and is difficult to measure in the offline setting. As a complement, some researchers \cite{202202} propose a multi-armed bandit-based algorithm to audit envy-free fairness based on true preferences.

\subsubsection{Metrics for Counterfactual Fairness (CF)}

Li et al. \cite{202115} demonstrate that counterfactual user fairness can be guaranteed when user embeddings are independent of fairness-related attributes. Therefore, they use a classifier to predict fairness-related attributes based on user embeddings and use classification measurements to measure counterfactual fairness. The classification measurements can be Precision, Recall, AUC, and F1 et al.

\subsubsection{Metrics for Rawlsian Maximin Fairness (RMF)}
Rawlsian maximin fairness argues that fairness depends on the worst individual or group. A simple measurement is the utility of the worst case, but it is vulnerable to noise. In order to make the metrics robust, some work \cite{202117,202122} uses the average utility of the bottom n\% as a measurement. The higher the value, the fairer the recommendations.

\subsubsection{Metrics for Maximin-shared Fairness (MSF)}
Maximin-shared fairness requires the outcome of each individual to be more than its maximin share. A measurement for item maximin-shared fairness is the proportion of individuals satisfying this condition, where the maximin share for every item is a constant value, i.e., the average exposure\cite{202009}. The higher the value, the fairer the recommendations.

\subsubsection{Metrics for Process Fairness (PR)}
One criterion of process fairness is that the model should use fair representations. A fair representation should be independent of fairness-related attributes, so some work \cite{202109,201904} trains a classifier to predict fairness-related attributes of users and items according to their representations. Then they use some classification measurements (e.g., precision) to measure the fairness of representations, which are similar to the counterfactual fairness measurements \cite{202115}.

\section{Methods for Fair Recommendation}
\subsection{Overview of Fairness Methods}
To our knowledge, existing methods for improving fairness can be categorized into three classes according to their working position in the recommendation pipeline, i.e., data-oriented methods, ranking methods, and re-ranking methods, as shown in Fig.\ref{fig:methods}.
Data-oriented methods are proposed to alleviate the unfairness problem by changing the training data. 
Ranking methods mainly design fairness-aware recommendation models or optimization targets for learning fair recommendations.
Re-ranking methods mainly adjust the outputs of recommendation models to improve fairness. 
Since there are more methods in the last two categories, we further grouped the methods in these two categories, and the specific sub-groups are also illustrated in Fig.\ref{fig:methods}.

{\begin{figure}[h]
  \centering
  \includegraphics[width=0.7\linewidth]{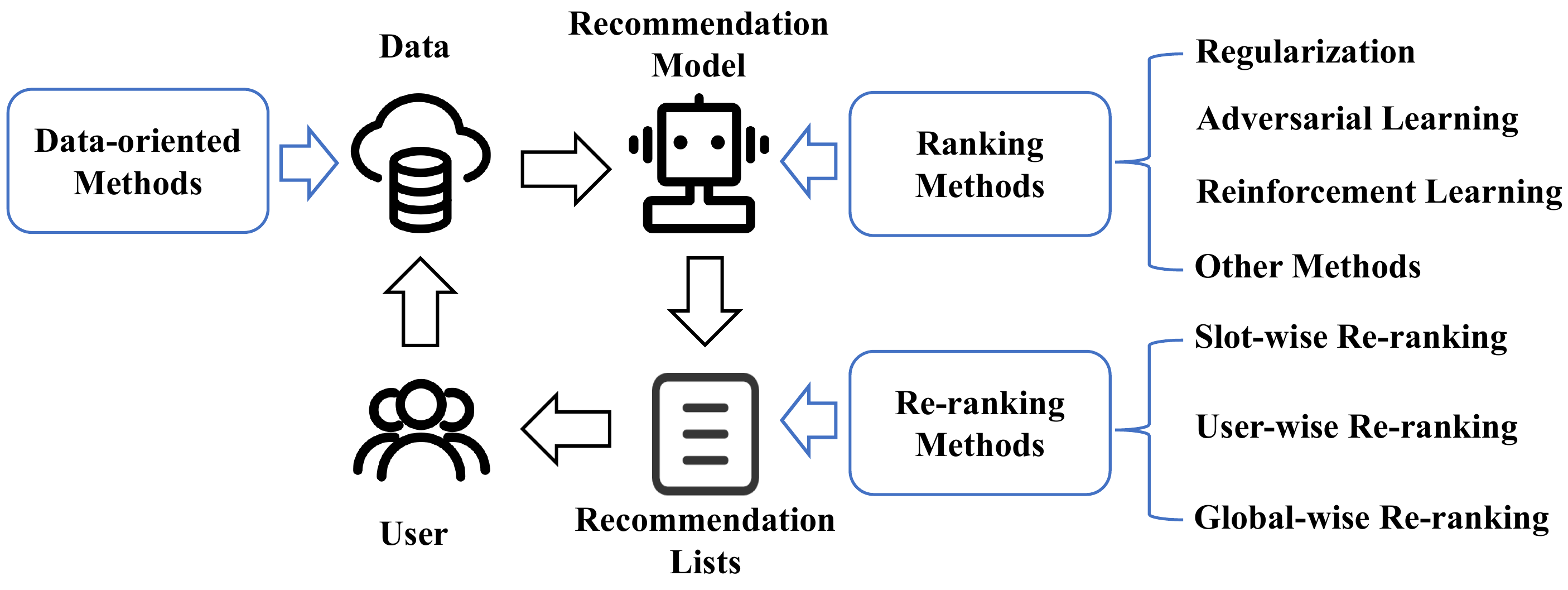}
  \caption{Taxonomy of fairness methods in the recommendation and their position in the recommendation pipeline.}
  \label{fig:methods}
  \Description{The recommendation pipeline consists of a directed loop from the user to the data, to the recommendation model, to the recommendation list, and back to the user. The data-oriented methods work in the data part, the ranking methods work in the recommendation model part, and the re-ranking methods work in the recommendation list part.}
\end{figure}}

The reviewed methods and corresponding brief descriptions are summarized in Table \ref{tab:methods}. It can be observed that there are only a few data-oriented methods. For ranking methods, regularization and adversarial learning are the dominant methods. At the same time, reinforcement learning has also gained attention in recent years due to being more suitable for modeling dynamics and long-term effects. For re-ranking methods, slot-wise re-ranking methods are dominant, but an increasing amount of recent work has focused on global-wise re-ranking.

\begin{table}[h]
\caption{A lookup table for the reviewed fairness methods in Recommendation.}
\label{tab:methods}
\centering
\resizebox{\textwidth}{!}{%
\begin{tabular}{l|l|l|l|l} 
\toprule \toprule
\textbf{Paper  }                        & \textbf{Type} & \textbf{Brief Description}           & \textbf{Publication}     & \textbf{Year}  \\[0.1em]
\midrule\midrule
\cite{201813} & data-oriented &  adjust the proportion of the protected group by resampling        & FAT*            & 2018  \\ 
\hline
\cite{201907} & data-oriented & add antidote data to the training data        & WSDM            & 2019  \\ 
\hline
\cite{201702} & regularization & use fairness metrics (e.g., value fairness) as fair regularization       & NIPS            & 2017  \\ 
\hline
\cite{201814} & regularization & use distribution matching and mutual information terms as regularization         & FAT*            & 2018  \\ 
\hline
\cite{201809} & regularization &  add fairness regularization to SLIM      & FAT*            & 2018  \\ 
\hline
\cite{201811} & regularization & induce orthogonality between insensitive latent factors and sensitive factors  & CIKM            & 2018  \\
\hline
\cite{201906} & regularization & add pairwise fairness regularization based on randomized experiments        & KDD             & 2019  \\ 
\hline
\cite{202008} & regularization &  use F-statistic of ANOVA  as regularization      & WSDM            & 2020  \\ 
\hline
\cite{201904} & adversarial learning & fairness constraints for graph embeddings   & ICML            & 2019  \\
\hline
\cite{202017} & adversarial learning & learn fair predicted scores by enhancing score distribution similarity  & SIGIR           & 2020  \\ 
\hline
\cite{202109} & adversarial learning & learn fair representations in graph-based recommendation  & WWW             & 2021  \\
\hline
\cite{202105} & adversarial learning & add text-based reconstruction loss to learn fair representations       & WSDM            & 2021  \\
\hline
\cite{202115} & adversarial learning & learn personalized counterfactual fair user representations & SIGIR           & 2021  \\ 
\hline
\cite{202106} & adversarial learning & learn fair user representation in news recommendation  & AAAI            & 2021  \\ 
\hline
\cite{202203} & adversarial learning & a GAN-based fair learning algorithm   & WWW           & 2022  \\ 
\hline
\cite{202011} & reinforcement learning & add fairness-related rewards to improve long term fairness     & PAKDD           & 2020  \\ 
\hline
\cite{202102} & reinforcement learning & add fairness-related constraints to improve long term fairness    & WSDM            & 2021  \\ 
\hline
\cite{202204} & reinforcement learning & achieve Pareto efficient fairness-utility tradeoff by multi-objective RL    & WSDM            & 2022  \\ 
\hline
\cite{201801} & other ranking method & a hybrid fair model with probabilistic soft logic  & RECSYS          & 2018  \\ 
\hline
\cite{201912} & other ranking method & add a noise component to VAE  & MEDES           & 2019  \\ 
\hline
\cite{202107} & other ranking method & use a pre-training and fine-tuning approach with bias correction techniques & WWW             & 2021  \\ 
\hline
\cite{202201} & other ranking method & adjust the gradient based on the predefined fair distribution & BIGDATARES.             & 2022  \\ 
\hline
\cite{201707} & slot-wise re-ranking & maximize ranking utility with group fairness constraint by two queues  & CIKM            & 2017  \\ 
\hline
\cite{201703} & slot-wise re-ranking & use greedy algorithm to maximize fairness in group recommendation  & WWW             & 2017  \\ 
\hline
\cite{201805} & slot-wise re-ranking & fairness-aware variation of the maximal marginal relevance  & UMAP            & 2018  \\ 
\hline
\cite{201816} & slot-wise re-ranking & calibrated recommendation through maximal marginal relevance & RECSYS          & 2018  \\ 
\hline
\cite{201905} & slot-wise re-ranking & improve multiple group fairness by interval constrained sorting & KDD             & 2019  \\ 
\hline
\cite{201913} & slot-wise re-ranking & find pareto optimal items in group recommendation  & SAC             & 2019  \\ 
\hline
\cite{201908} & slot-wise re-ranking & personalized fairness-aware re-ranking   & RECSYS          & 2019  \\ 
\hline
\cite{202005} & slot-wise re-ranking & personalized fairness-aware re-ranking with different user tolerance   & UMAP            & 2020  \\ 
\hline
\cite{202010} & slot-wise re-ranking & ensure fairness in group recommendation in a ranking sensitive way   & RECSYS          & 2020  \\
\hline
\cite{202014} & slot-wise re-ranking & ensure fairness in dynamic learning to rank through p-controller  & SIGIR           & 2020  \\ 
\hline
\cite{202110} & slot-wise re-ranking & ensure fairness in dynamic learning to rank by maximal marginal relevance   & WWW             & 2021  \\ 
\hline
\cite{202206} & slot-wise re-ranking & enumerate fair packages for group recommendations  & WSDM             & 2022  \\ 
\hline
\cite{201704} & user-wise re-ranking & fair group recommendation from the perspective of Pareto Efficiency  & RECSYS          & 2017  \\ 
\hline
\cite{201810} & user-wise re-ranking & a series of recommendation policies to combine fairness and relevance   & CIKM            & 2018  \\ 
\hline
\cite{201808} & user-wise re-ranking & ensure amortized fairness through integer linear programming  & SIGIR           & 2018  \\ 
\hline
\cite{201819} & user-wise re-ranking & linear programming from the perspective of probabilistic rankings  & KDD             & 2018  \\ 
\hline
\cite{202205} & user-wise re-ranking & mitigate outlierness in fair rankings through linear programming  & WSDM             & 2022  \\ 
\hline
\cite{201817} & global-wise re-ranking & 0-1 integer programming with providers constraint & RECSYS          & 2018  \\ 
\hline
\cite{202009} & global-wise re-ranking & a re-ranking method for both user fairness and item fairness & WWW             & 2020  \\ 
\hline
\cite{202003} & global-wise re-ranking & fairness-aware explainable recommendation through 0-1 integer programming & SIGIR           & 2020  \\ 
\hline
\cite{202121} & global-wise re-ranking & a re-ranking method based on maximum flow & TOIS            & 2021  \\ 
\hline
\cite{202112} & global-wise re-ranking & ensure user group fairness through 0-1 integer programming & WWW             & 2021  \\ 
\hline
\cite{202122} & global-wise re-ranking & a re-ranking method for joint fairness via Lorenz dominance & NIPS             & 2021  \\ 
\hline
\cite{202114} & global-wise re-ranking & a re-ranking method for both user fairness and provider fairness & SIGIR           & 2021  \\ 
\hline
\cite{202117} & global-wise re-ranking & a learnable re-ranking method for fairness among new items & SIGIR           & 2021  \\ 
\bottomrule \bottomrule
\end{tabular}
}
\end{table}

As shown in Table \ref{tab:types}, we also summarize the types of fairness issues solved by each method type. It can be found that each method type can solve several different types of fairness issues, and most fairness issues, on the other hand, can be solved using multiple methods. However, some fairness issues are more specific. For example, process fairness and counterfactual fairness issues are solved only using adversarial learning. Rawlsian maximin and maximin-shared fairness tend to be solved using global-wise re-ranking. Indeed, it may be because there is less work related to these fairness issues. It is also worth exploring to design other methods to solve these issues.

\begin{table}[h]
\small
\caption{The current types of fairness issues solved by each method type. (Here "CO" means consistent fairness, "CA" means calibrated fairness, "CF" means counterfactual fairness, "EF" means envy-free fairness, "RMF" means Rawlsian maximin fairness, "PR" means process fairness and "MSF" means maximin-shared fairness).}
\label{tab:types}
\centering
\resizebox{\textwidth}{!}{%
\begin{tabular}{l|l|p{1cm}<{\centering}c|ccc|cc|cc}
\toprule \toprule
\multirow{2}{*}{\textbf{Method Type}}   & \multirow{2}{*}{\textbf{Def.}}                                                     & \multicolumn{2}{c|}{\textbf{Target}}                                            & \multicolumn{3}{c|}{\textbf{Subject}} &
\multicolumn{2}{c|}{\textbf{Granularity}} &  \multicolumn{2}{c}{\textbf{Optim. Object}}\\[0.1em]
\cline{3-11} \rule{0em}{2.5ex}
    &   & Group & Ind. & User    & Item & Joint   & Single & Amortized  & Treat. & Impact                              \\ 
\midrule \midrule
\bigcell{l}{Data-oriented \\ \cite{201813,201907}} & CO & \checkmark & \checkmark & \checkmark & \checkmark & & & \checkmark & & \checkmark \\ \hline
\bigcell{l}{Regularization \\ \cite{201702,201814,201809,201811,201906,202008}} & CO \& CA & \checkmark & & \checkmark & \checkmark & \checkmark  & \checkmark & \checkmark & \checkmark & \checkmark \\
\bigcell{l}{Adversarial Learning \\ \cite{201904,202017,202109,202105,202115,202106,202203}} & PR \& CO \& CF & \checkmark & \checkmark &\checkmark &\checkmark & & \checkmark & \checkmark & \checkmark & \checkmark \\
\bigcell{l}{Reinforcement Learning \\ \cite{202011,202102,202204}} & CO \& CA & \checkmark & & & \checkmark & & & \checkmark & \checkmark & \\
\bigcell{l}{Others \\ \cite{201801,201912,202107,202201}} & CO \& CA & \checkmark & \checkmark & \checkmark & \checkmark & & & \checkmark & \checkmark & \checkmark \\ \hline
\bigcell{l}{Slot-wise Re-ranking \\ \cite{201707,201703,201805,201816,201905,201913,201908} \\ \cite{202005,202010,202014,202110,202206}} & CO \& CA \& EF & \checkmark & \checkmark & \checkmark & \checkmark & & \checkmark & \checkmark & \checkmark & \checkmark \\
\bigcell{l}{User-wise Re-ranking \\ \cite{201704,201810,201808,201819,202205}} & CO \& CA & \checkmark & \checkmark & \checkmark & \checkmark & & \checkmark & \checkmark & \checkmark & \checkmark \\
\bigcell{l}{Global-wise Re-ranking \\ \cite{201817,202009,202003,202121,202112,202114,202117,202122}} & \bigcell{l}{CO \& CA \& EF \\ \& MSF \& RMF} & \checkmark & \checkmark& \checkmark& \checkmark& \checkmark& & \checkmark& \checkmark& \checkmark\\
\bottomrule \bottomrule
\end{tabular}%
}
\end{table}

\subsection{Data-oriented Methods}
The data-oriented methods improve fairness by modifying the training data. Compared with other types of methods, there are fewer data-oriented methods.

Considering that user unfairness might result from the data imbalance between different user groups, Ekstrand et al. \cite{201813} use re-sampling to adjust the proportion of different user groups in the training data. Experiments on the Movielens 1M dataset show that this approach can alleviate unfairness, but not significantly. 

Rastegarpanah et al. \cite{201907} design a relatively more complex but effective method. They draw on data poisoning attacks to address the unfairness problem by adding additional antidote data (e.g., fake user data) to the training data. Adding antidote data during training will affect the predicted rating matrix, which further affects the fairness of recommendations. The antidote data can be updated by optimizing the fairness objective function through the gradient descent method. Compared to the re-sampling method, this approach can better mitigate unfairness, but it is also relatively more time-consuming.

In summary, we can adjust the training data to improve the fairness of recommendations. The advantage of these methods is their low coupling with the recommender system since these methods do not require modification of the original recommendation model. Besides, as these methods work at the front part of the recommendation pipeline, there are fewer constraints on the candidate set. They have the potential to improve the fairness of the recommendation results significantly. However, since multiple stages exist between the data and the final presentation, their performance might be degraded by subsequent stages such as re-ranking for diversity. It is challenging to design effective data-oriented methods.

\subsection{Ranking Methods}
Ranking methods mainly modify recommendation models or optimization targets to learn fair representations or prediction scores. The ranking is the main focus of research in recommendation techniques. It is natural to use some advanced techniques to solve the problems of fair representation learning and long-term fairness, which is difficult for the other two types of methods. Compared to data methods, the results of sorting methods are less different from the final presentation, and the improvement in fairness is more straightforward. Nevertheless, since a re-ranking stage may exist after the ranking stage, similar to data-oriented methods, their performance may be damaged by downstream re-ranking stages.

Depending on the different techniques, current fairness methods for the ranking phase can be divided into regularization-based methods, adversarial learning-based methods, reinforcement learning-based methods, and others. 

\subsubsection{Regularization}
One common approach is adding a fairness-related regularization term to the loss function. Formally, denote $L_{rec}$ as the traditional recommendation loss function and $L_{fair}$ as the fairness-related regularization term, then the loss function considering fairness is formalized as $L = L_{rec} + \lambda \cdot L_{fair}$.

\smallskip
One \textbf{direct} approach is to add the fairness evaluation metrics \cite{201702,201814,202008,202015} to the loss function as a regularization term, which requires that the metric is differential. It is difficult to use this approach to address unfairness in exposure or ranking as the corresponding metrics are not differential, so existing related work is more focused on unfairness in rating prediction. The advantage of this approach is its simplicity and effectiveness, while the disadvantage is that it is limited in application and often results in a loss of recommendation performance. 

\smallskip
In contrast, some approaches \cite{201811,201809,201906} impose \textbf{indirect} regularization on the model. Compared to direct methods, indirect methods can achieve better fairness and recommendation performance. Here we introduce some representative methods below.

In order to reduce the correlation between predicted scores and fairness-related attributes, Zhu et al. \cite{201811} propose a fairness-aware tensor-based recommendation framework(FATR), which induces orthogonality between the representations of users (or items) and the corresponding vector of fairness-related attributes by adding a regular term in the tensor-based recommendation model. The loss function is the following Eq.(\ref{eq:201811}) \cite{201811}. The first term of the loss function is the original part of the tensor-based recommendation model. The second term is the regular fairness term, which extracts fairness-related attribute information in latent factor matrices. The final fairness prediction is calculated as $[\![ A_1',...,A_n',...,A_N' ]\!]$. 
\begin{equation}
    \begin{gathered}
    \mathop{minimize}\limits_{X,A_1,...,A_n',...,A_N} L = ||X - [\![ A_1,...,\tilde{A_n},...,A_N ]\!]||_F^2 + \frac{\lambda}{2}||A_n^{''T}A_n^{'}||_F^2 + \frac{\gamma}{2}\sum_{i = 1}^N||A_i||_F^2 \\
    s.t. \quad \Omega \circledast X = J,\tilde{A_n} = [A_n' \ A_n''], A_n'' = S\label{eq:201811}
    \end{gathered}
\end{equation}
Here $X$ is a tensor denoting the complete preferences of users, $[\![ \cdot ]\!]$ is the Kruskal operator, $[A \ B]$ is the matrices concatenating operator, $\odot$ is the Khatri-Rao product, and $\circledast$ is the Hadamard product. $J$ denotes observations, and $\Omega$ is the non-negative indicator tensor indicating whether we observe $X$. $A_1,...,A_N$ denote the latent factor matrices of all the modes of the tensor. Here $A_n \in R^{d_n \times r}$ is the latent factor matrix of the fairness-related mode mode-n, where $r$ is the dimension of the latent factors and $d_n$ is the entity number of the mode-n, and it can be split into two part $A_n'$, $A_n''$. 

Experiments on real datasets show that FATR can achieve better recommendation performance and fairness than directly using the fairness metric as a regular term, reflecting the advantages of the indirect approach.

\smallskip
While the above work focuses on the fairness of point-wise predicted scores, Beutal et al. \cite{201906} investigate the fairness from the perspective of pair-wise ranking. They demonstrate that the fairness of point-wise ranking tasks does not guarantee the fairness of pair-wise ranking. To improve pair-wise ranking fairness, they add the residual correlations of fairness-related attributes and predicted preferences as regular terms to motivate the model to have similar prediction accuracy across item groups. The loss function is the following Eq.(\ref{eq:201906}) \cite{201906}. The second term is the regular fairness term, which will be bigger if the model has a better prediction ability for the clicked item in one group than the other.
\begin{equation}
\begin{gathered}
\mathop{min}\limits_\theta(\sum\limits_{(q,j,y,z) \in D} L(f_\theta(q,v_j),(y,z))) + |Corr_P(A,B)| \\
A = (g(f_\theta(q,v_j)) - g(f_\theta(q,v_j')))(y - y') \\
B = (s_j - s_j')(y - y')\label{eq:201906}
\end{gathered}
\end{equation}
Here $s_j$ is the binary fairness-related attribute for item $j$, $q$ is the query consisting of user and context features, $y$ is the user click feedback, $z$ is the post-click engagement, $f_\theta(q,v)$ is the predictions $(\hat{y},\hat{z})$ for item $v$, $g(\hat{y},\hat{z})$ is the monotonic ranking function from predictions. $P$ is experimental data, and both $A$ and $B$ are random variables over pairs from $P$. 

\smallskip
In summary, we can add a fairness-related regularization term to the loss to improve fairness. Compared to other ranking methods, regularization-based methods are more flexible and easily extensible. However, simply adding regularization terms may make it difficult for the model to learn fairness-related information, which might lead to suboptimal performance.

\subsubsection{Adversarial Learning}
\begin{figure}[htbp]
\centering
\subfigure[Learning fair representations]{
\begin{minipage}[t]{8.9cm}
\centering
\includegraphics[scale=0.4]{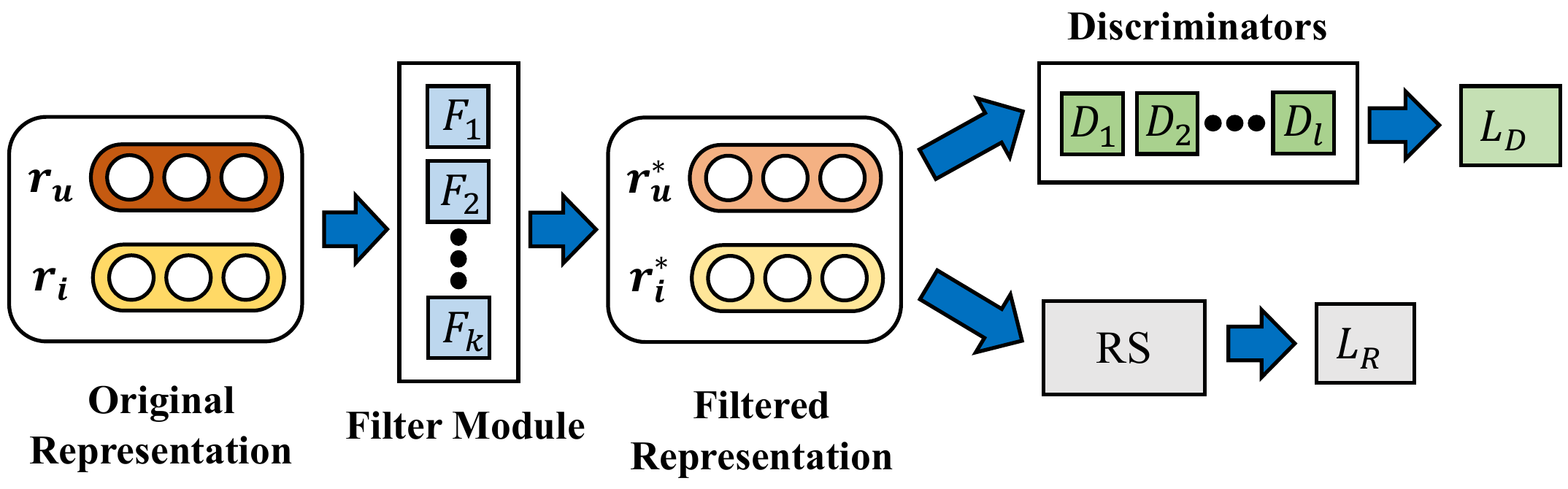}
\end{minipage}
}
\subfigure[Learning fair predicted scores]{
\begin{minipage}[t]{4.9cm}
\centering
\includegraphics[scale=0.42]{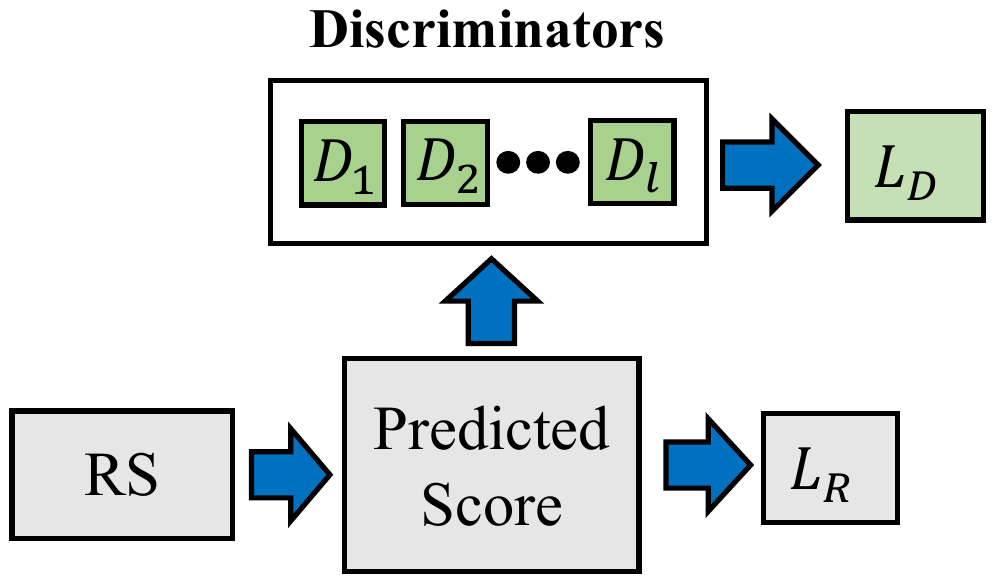}
\end{minipage}
}
\caption{Basic frameworks of adversarial learning.}
\label{fig:adversariallearning}
\Description{The figure is fully described in the text.}
\end{figure}

Several studies use adversarial learning to address the fairness problem \cite{201904,202106,202109,202017,202115,202105}. As mentioned earlier, process fairness requires that recommender systems use fair representations. Even though sensitive information is not directly used as input, it may still be indirectly learned by the model into the representation. Adversarial learning is an effective method to reduce the sensitive information in the representation. In addition, it can also be applied to learn fair predicted scores. The basic frameworks of adversarial learning are illustrated in Fig.\ref{fig:adversariallearning}.

\smallskip
A series of studies \cite{201904,202109,202106,202115} are aimed to learn fair representation through adversarial learning. The basic framework is shown in Fig.\ref{fig:adversariallearning}(a). Apart from the recommendation model, they often introduce a discriminator for each fairness-related attribute. These discriminators will predict the corresponding attribute value based on the representations outputted by a filter module which is designed to remove unfair information in original representations. If the discriminator cannot determine the value of these fairness-related attributes according to the filtered representations, the filtered representations will be fair.
The learning process can be formalized as the following two-player minimax game.
\begin{equation}
\mathop{min}\limits_{R} \mathop{max}\limits_{D} L(R,D) = L_R - \lambda L_D 
\end{equation}
Here $L_R$ is the recommendation loss, and $L_D$ is the attribute prediction loss of discriminators. $R$ is the parameters for the recommendation model, and $D$ is the parameters of discriminators. $\lambda$ is a hyperparameter. We briefly introduce these methods below.

Avishek and William \cite{201904} propose a method to reduce the sensitive information contained in the node representation in graph neural networks, which can be applied to multiple fairness-related attributes simultaneously. This method introduces multiple filter modules in the model, each corresponding to a fairness-related attribute, and removes the corresponding fairness-related attribute information from the node representation. After sensitive information is filtered, all filtered representations of that node are averaged together to obtain a representation without sensitive information $v$. The discriminator of each fairness-related attribute will predict the corresponding fairness-related attribute of that node based on the representation $v$. For recommendation, they will only use fair representations $v$.

In addition to node representations, the network structure around nodes is also important information, which is ignored in the above approach. Wu et al. \cite{202109} add discriminators to the graph network recommendation model, which predicts the fairness-related attributes of nodes based on their embeddings and the embeddings of the network structure around the nodes. Experimental results on real datasets also validate that better results can be achieved than the method considering only node information.

The discriminator proposed by Li et al. \cite{202115} also predicts the fairness-related attributes of users based on their embeddings. The main difference from previous work is that users can personalize their fairness-related attribute settings.

Unlike the above methods, Wu et al. \cite{202106} focus on the fairness of user representations in news recommendations, where the user representation is constructed from the user's reading history. They add a discriminator to the news recommendation model to learn fair user embeddings, which predicts the fairness-related attributes of users based on their embeddings. Besides, they also add an attribute prediction task to learn unfair user embeddings. Furthermore, they add regularization to the loss function to enhance the orthogonality between fair and unfair embeddings.

\smallskip
Apart from learning fair representations, adversarial learning can also be used to learn fair predicted scores. Zhu et al. \cite{202017} add discriminator to the recommendation model, which predicts the fairness-related attributes of items based on the predicted scores of the recommendation model. Then they ensure item fairness through the adversary between the recommendation model and the discriminator. The training process can be formalized as the following Eq.(\ref{eq:202017}) \cite{202017}.
\begin{equation}
\mathop{min}\limits_\Theta \mathop{max}\limits_\Psi \sum_{u \in U} \sum_{i \in I_u^+, j \in I\backslash I_u^+} (L_{BPR}(u,i,j) + \alpha (L_{Adv}(i) + L_{Adv}(j))) + \beta L_{KL} \label{eq:202017}
\end{equation}
Here $L_{Adv}(i)$ is the log-likelihood loss for an MLP adversary to classify items, and $L_{KL}$ is the KL-loss between the score distribution of each user and a standard normal distribution, which will make the score distribution of each user conform to normal distribution. $L_{BPR}$ is the recommendation loss. $\Theta$ and $\Psi$ are learnable parameters for the recommendation model and discriminator. $I_u^+$ is the set of positive items for user $u$. $\alpha$ and $\beta$ are hyperparameters.

\smallskip
Unlike the above methods that use discriminators to predict fairness-related attributes, some methods \cite{202105,202203} utilize discriminators in other ways. Li et al. \cite{202105} apply discriminators to reconstruct user/item information. They leverage textual information to improve user fairness. The key point is to promote user and item representations to restore the original textual information as much as possible and thus reduce the mainstream bias in minority representations. In addition, Li et al. \cite{202203} propose a GAN-based algorithm consisting of a ranker and a controller. Each component contains both a discriminator and a generator. The ranker learns user preferences, and its discriminator distinguishes real interactions from model-generated interactions. The controller provides fairness signals to make the ranker fair, and its discriminator distinguishes the generated exposure distribution from the exposure distribution calculated based on the predictions of the ranker.

\smallskip
In summary, current work often leverages adversarial learning to learn fair representations or predicted scores to improve recommendation fairness. Adversarial learning is well-suited to fair representation learning and is the dominant approach to this problem. However, since its optimization objective is a minimax optimization problem, it is more difficult to train than the traditional minimization problem.

\subsubsection{Reinforcement Learning}
\begin{figure}[htbp]
\centering
\subfigure[RL for recommendation]{
\begin{minipage}[t]{6.9cm}
\centering
\includegraphics[scale=0.4]{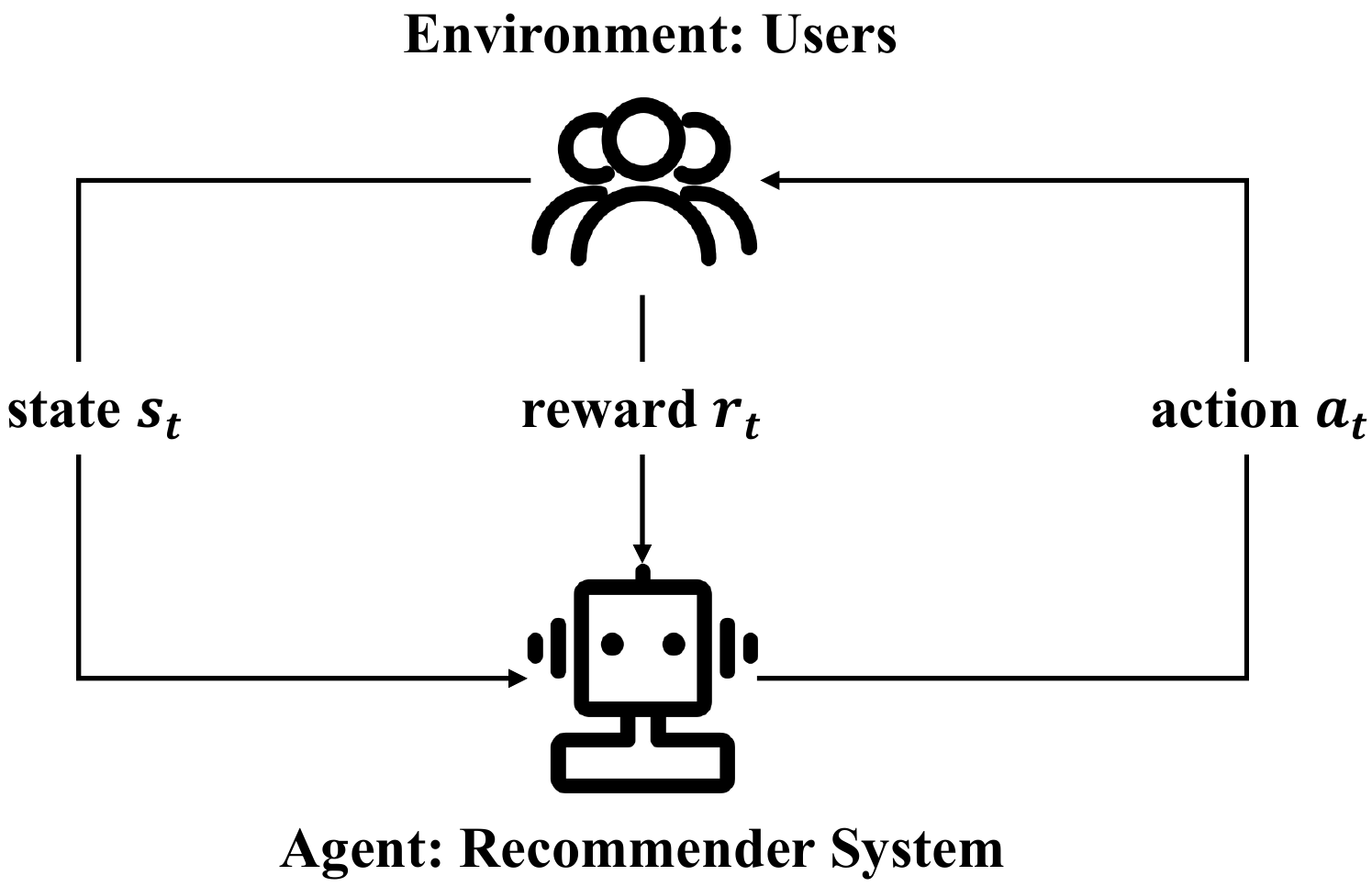}
\end{minipage}
}
\subfigure[RL for the fair recommendation]{
\begin{minipage}[t]{6.9cm}
\centering
\includegraphics[scale=0.4]{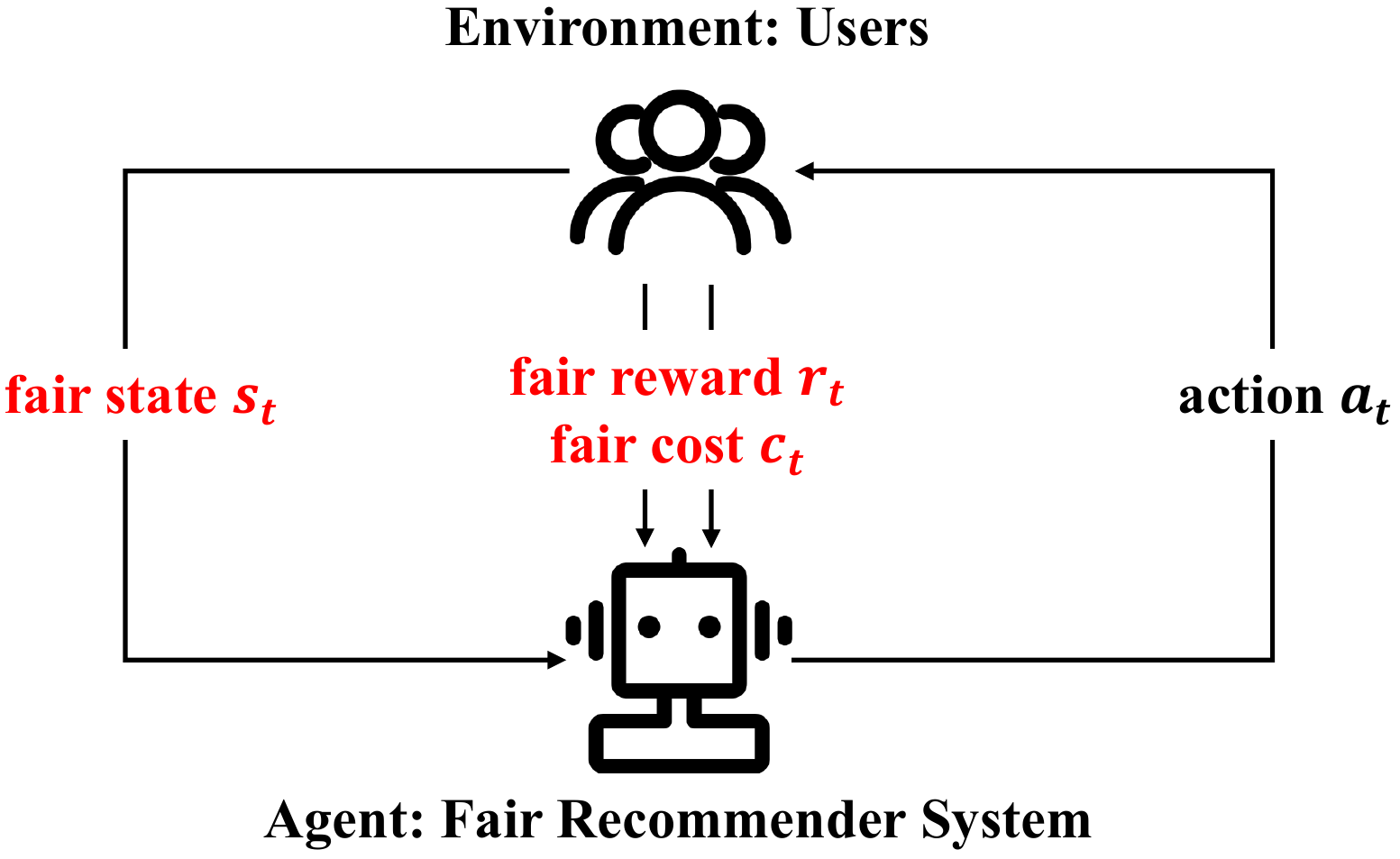}
\end{minipage}
}
\caption{Illustrations of RL for recommendation and RL for the fair recommendation. Note that (b) shows all the differences between the existing fair RL methods and the general RL methods for the recommendation. In other words, certain fair RL methods may not contain all the differences.}
\label{fig:reinforcementlearning}
\Description{The figure is fully described in the text.}
\end{figure}

Several studies use reinforcement learning (RL) to address the fairness problem \cite{202102,202011}. Compared to other methods which mainly consider the immediate fairness impact, reinforcement learning-based fairness methods can optimize fairness in the long run. Fig.\ref{fig:reinforcementlearning} shows all the differences between existing fair RL methods and general RL methods for the recommendation. Current work mitigates unfairness in recommendations by introducing fairness information in states, rewards, or additional constraints.

An intuitive way to improve fairness is to design fairness-related rewards. In order to improve long-term fairness, Liu et al. \cite{202011} first propose a reinforcement learning-based method. They introduce fairness-related rewards to make recommendations fair. The reward is defined as the following Eq.(\ref{eq:202011}) \cite{202011}.
\begin{equation}
r_t=
\begin{cases}
\sum_{i = 1}^l 1_{A_{c_i}}(a_t)(x_*^i - x_t^i + 1), & if \  y_{a_t} = 1\\
- \lambda, & if \ y_{a_t} = 0 \label{eq:202011}
\end{cases} 
\end{equation}
Here $x_*^i$ is the optimal allocation for group $i$ and $x_t^i$ is the allocation for group $i$ in time step $t$. $A$ is the item set and $A_{c_i}$ is the item with the attribute value $c_i$. $y_{a_t}$ is the user feedback on item $a_t$. $\lambda$ is a hyperparameter.

They also propose a reinforcement learning-based model based on the actor-critic architecture. The actor-network learns a dynamic fairness-aware ranking strategy vector $z$, which contains user preferences and the system's fairness status. Then ranking score is calculated based on $z$ and item ID embedding. The critic-network estimates the value according to $z$ and a fairness allocation vector, which provides information about the current allocation distribution of different groups.

Similarly, Ge et al. \cite{202204} also improve fairness by introducing fairness-related rewards. Still, the difference is that they formalize the problem as a multi-objective Markov decision process and solve it using conditioned networks. Their approach can seek the Pareto frontier of fairness and utility, thus facilitating decision-makers to control the fairness-utility trade-off.

\smallskip
The above work adds fairness signals by changing the rewards, while some work promotes fairness by adding fairness-related constraints. Ge et al. \cite{202102} consider the dynamics in long-term fairness, in other words, the changes of group labels or item attributes due to the user feedback during the whole recommendation process. The dynamic fairness problem is modeled as the constrained Markov decision process, which has been well studied. Specifically, they add constraints to ensure the fairness of recommendations. The constraint is defined as the following Eq.(\ref{eq:202102}).
\begin{equation}
    \frac{Exposure_{t}(G_0)}{Exposure_{t}(G_1)} \le \alpha \label{eq:202102}
\end{equation}
Here $Exposure_{t}(G_0)$ and $Exposure_{t}(G_1)$ are the number of exposure in group $G_0$ and group $G_1$ at iteration $t$, and $\alpha$ is a hyperparameter.

They additionally define the cost function as the number of sensitive group items in the recommendation list and find that the fairness constraint can be transformed into a constraint on the cost function. Thus the fairness problem can be formalized as a Markov decision problem with constraints and then solved. They also apply the actor-critic architecture, but the main difference is that their model contains two critics, which approximate the reward and cost, respectively. Compared to the above method containing explicit input about fairness status \cite{202011}, this model has no fairness-related explicit input.

\smallskip
In summary, existing work on reinforcement learning achieves fair recommendations via modifications to the state, reward, or additional constraints. Compared to other methods, reinforcement learning can optimize long-term and dynamic fairness. Nevertheless, reinforcement learning is difficult to evaluate with offline data and has poor stability \cite{200062}.

\subsubsection{Other Methods}
There are also several other fairness ranking methods. Islam et al. \cite{202107} use transfer learning to learn fair user representations for career recommendations, and they propose a fair neural model based on neural collaborative filtering (NCF) \cite{200053}. They first learn a pre-trained model on insensitive items, then transform the pre-trained user embeddings to mitigate fairness-related attribute biases in them, and then fine-tune them on sensitive items. 

Li et al. \cite{202201} propose a contextual framework for the fairness-aware recommendation, which is suitable for different fair performance distributions. Specifically, the framework will infer a coefficient for each user/item from the predefined fair distribution. Then the framework will adjust the gradient during the optimization process based on the coefficient.

Borges et al. \cite{201912} improve recommendation fairness by adding a stochastic component to a trained VAE model. They find that introducing a normally distributed noise with high variance to the sampling phase can promote fairness despite a slight loss in the recommendation performance. 

Farnadi et al. \cite{201801} propose a rule-based fairness method. They use probabilistic soft logic to implement a fairness-aware hybrid recommender system.

\subsection{Re-ranking Methods}
Re-ranking methods mainly adjust the outputs of recommendation models to promote fairness. Re-ranking methods have the advantage that their results are nearly identical to the final presentation, making their improvement in outcome fairness the most straightforward. Besides, similar to data-oriented methods, they also have low coupling with the recommender system, as they do not require changing recommendation models. However, because the candidate set in the re-ranking stage is typically small, the performance of re-ranking methods may be hampered. Moreover, they cannot resolve the fair representation issue in the ranking stage.

\begin{figure}[htbp]
\centering
\subfigure[Slot-wise re-ranking]{
\begin{minipage}[t]{4.4cm}
\centering
\includegraphics[scale=0.45]{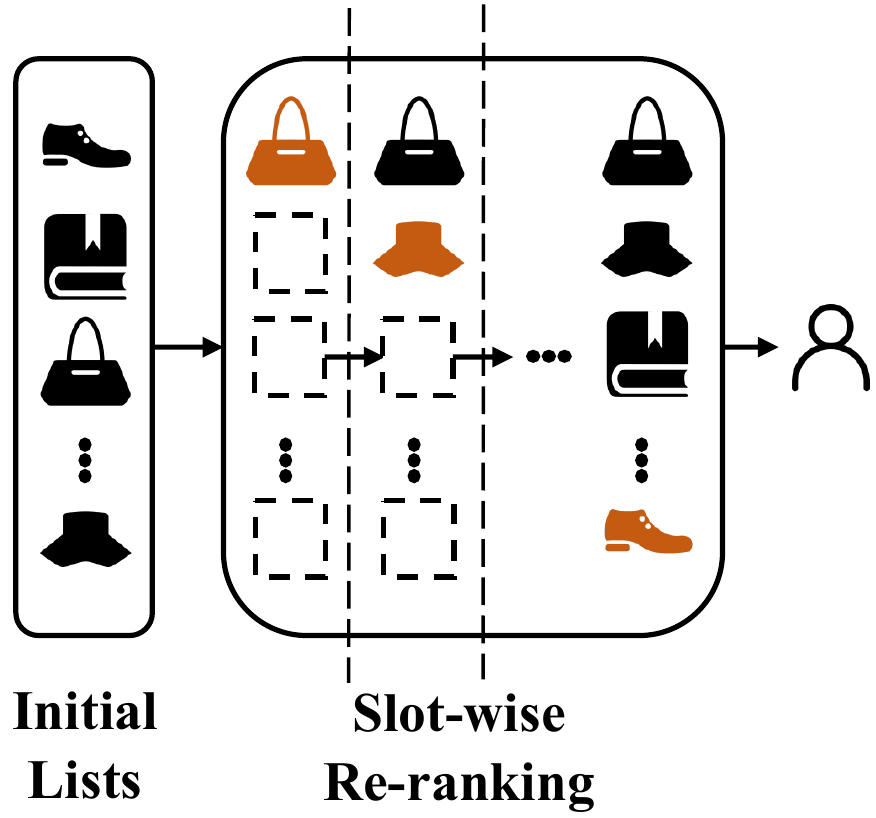}
\end{minipage}
}
\subfigure[User-wise re-ranking]{
\begin{minipage}[t]{4.4cm}
\centering
\includegraphics[scale=0.45]{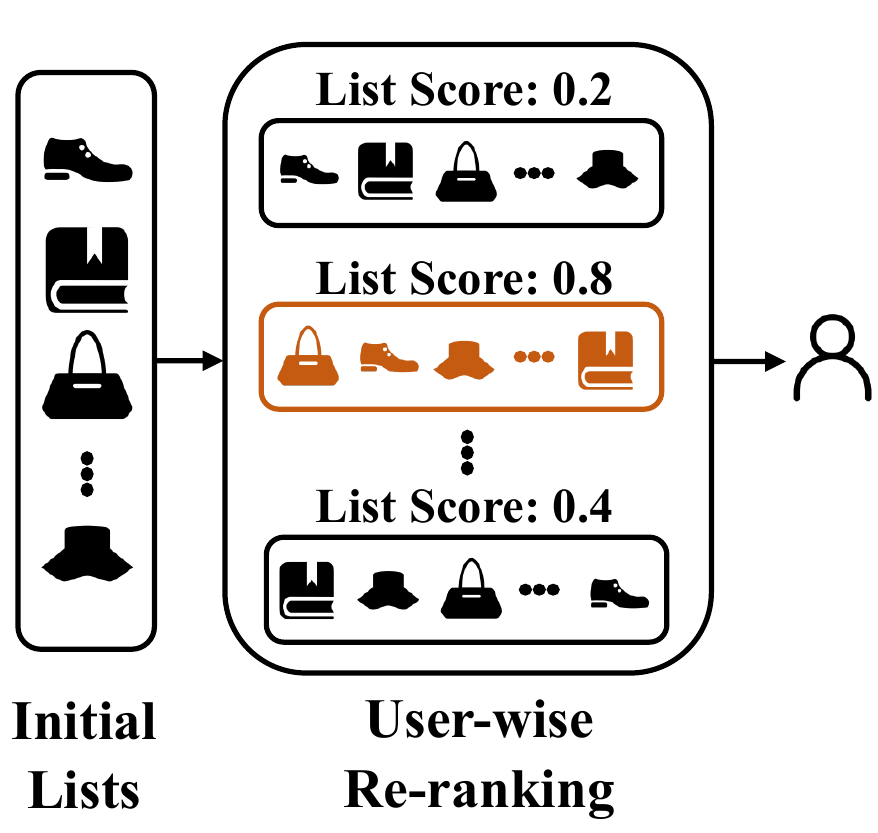}
\end{minipage}
}
\subfigure[Global-wise re-ranking]{
\begin{minipage}[t]{4.4cm}
\centering
\includegraphics[scale=0.45]{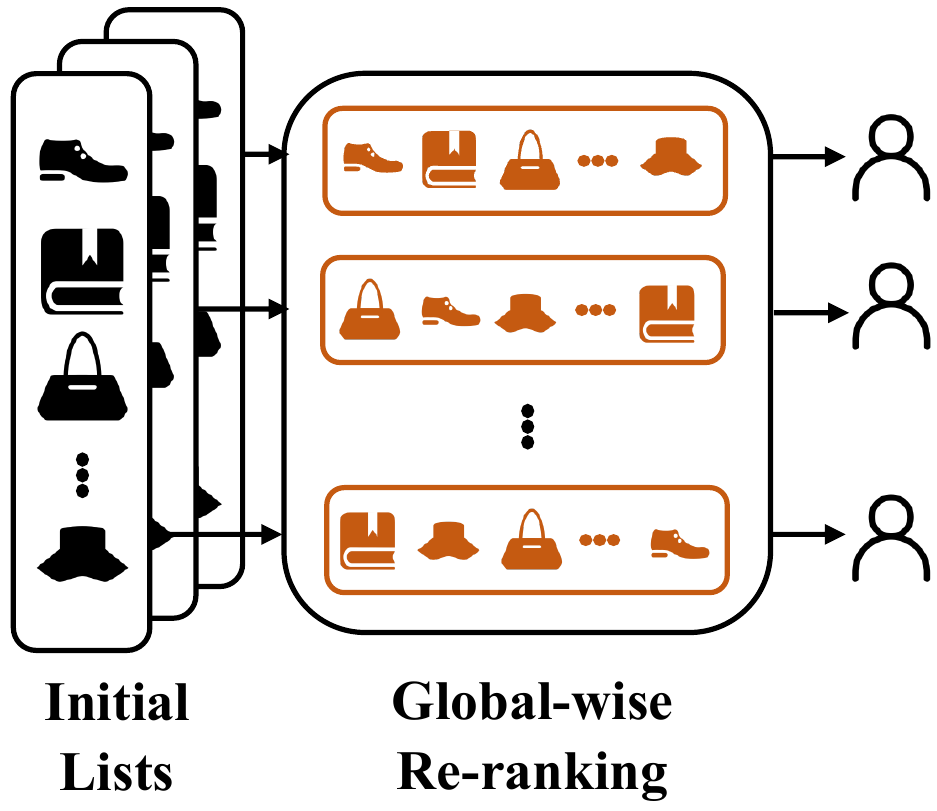}
\end{minipage}
}
\caption{Illustrations for different re-ranking types.}
\label{fig:reranking}
\Description{The figure is fully described in the text.}
\end{figure}

We divide current re-ranking methods into the following three types: \textbf{slot-wise}, \textbf{user-wise}, and \textbf{global-wise}. Fig.\ref{fig:reranking} illustrates the differences between these three types.
The slot-wise re-ranking method re-ranks a recommendation list by adding items to empty slots in a list one by one. It will select the next item added to the recommendation list by certain rules or re-ranking scores. 
Unlike slot-wise methods, user-wise re-ranking methods try to directly find the best recommendation list for a user based on the optimization goal of the whole list.
While the above two kinds of re-ranking methods are used for a single recommendation list each time, the global-wise re-ranking methods re-rank multiple recommendation lists for multiple users simultaneously. The re-ranking result for one recommendation list may be influenced by other lists.

\subsubsection{Slot-wise}
A few studies \cite{201704,201703,202010} propose slot-wise re-ranking methods to improve user fairness in \textbf{group recommendations}. Serbos et al. \cite{201703} use a greed-based approach to guarantee user fairness in group recommendations. They define the number of satisfied users for a recommendation package $P$ as $SAT_G(P)$. Then the gain from adding a new item $i$ to the current recommendation package $P$ can be defined as $f_G(P,i) = |SAT_G(P  \cup {i}) \backslash SAT_G(P)|$. The recommendation package can be constructed greedily, i.e., start with the empty set and gradually add items to the set to maximize $f_G(P, i)$. Lin et al. \cite{201704} also design a greedy algorithm to ensure user fairness in the group recommendation scenario. The difference is that they consider the Pareto efficiency between fairness and recommendation performance. They define the overall recommendation performance as $SW(g,I)$ and the fairness utility as $F(g,I)$. Then the Pareto frontier can be obtained by maximizing $\lambda \times SW(g,I) + (1 - \lambda) \times F(g,I)$. Further, recommendations can be obtained by adding items to the recommendation list one by one through a greedy strategy. Similarly, Sacharidis \cite{201913} finds Pareto optimal items to promote fairness. After obtaining the candidate Pareto optimal items, they generate ranking scores by linear aggregation strategies and estimate the probability of an item being ranked in Top-K in any strategy. Items are finally ranked based on the estimated probability. While previous studies considered only fixed-length recommendation lists, Kaya et al. \cite{202010} consider the fairness of different positions simultaneously. The method greedily selects each item and optimizes the following objective.
\begin{equation}
\begin{gathered}
i^* = \mathop{argmax}\limits_{i \in C \backslash OS} f(i,OS) \\
 f(i, OS) = f(OS \cup \{ i \}) - f(OS)\\
 f(S) = \sum_{u \in G} (1 - \prod_{i \in S} (1 - p(rel | u,i))) \\
\end{gathered}
\end{equation}
Here $C$ is the set of candidate items. $OS$ is the recommendation list recommended to the group. $p(rel|u,i)$ is the probability that item $i$ is relevant to user $u$.

\smallskip
In the \textbf{general recommendation} scenario, existing methods introduce fairness from two main perspectives, one is to maximize utility while satisfying fairness constraints, and another is to optimize both fairness and utility jointly. The former class of methods ensures that the final result is compliant with the fairness constraint, which may entail a relatively large performance loss, while the latter class of methods makes a trade-off between performance and fairness.


Some studies \cite{201707,201905} propose algorithms to satisfy the fairness constraint as much as possible at each position. Zehlike et al. \cite{201707} propose a priority queue-based approach FA*IR for item fairness scenarios where only two groups exist. FA*IR will maintain two priority queues sorted by relevance, corresponding to two groups. At each position, FA*IR will determine whether the current representation of the protected group satisfies the fairness constraint. If not, select the item with the highest relevance in the protected queue; otherwise, compare the two queues and select the item with the highest relevance. Based on FA*IR, Geyik et al. \cite{201905} propose three slot-wise methods to re-rank the results for more than three groups. The first two methods can be considered as extensions of FA*IR. The third algorithm regards fairness constrained re-ranking as an interval constrained sorting problem.

To make a trade-off between fairness and recommendation performance in the re-ranking phase, several studies \cite{201805,201816,201908,202005} optimize fairness and utility jointly, and provide hyperparameters to control the loss of recommendation performance. The process of these algorithms can be formalized as the following equation.
\begin{equation}
    i^* = \mathop{argmax}\limits_{i \in R \backslash C} \lambda P(u,i) + (1 - \lambda)F(u,i,C)
\end{equation}
Here $R$ is the set of item candidates. $C$ is the recommendation list for user $u$, which is empty initially. $P(u,i)$ is the predicted preference of user $u$ to item $i$. $F(u,i,C)$ is the fairness score. $\lambda$ is a hyperparameter to control the trade-off between fairness and utility. For each time, these algorithms will select an item $i^*$ from all the available candidate items and then put it into the recommendation list $C$. 

Different studies have different definitions of fairness score $F(u,i,C)$, which is the main difference between them. Steck \cite{201816} defines the fairness score from the user perspective and argues that the recommendation should be calibrated for the interests of users, which are measured through interaction history. The fairness score is defined as the KL-divergence between the distribution over different item groups in the history of user $u$ and the distribution over item groups in the recommendation list $C \cup \{i\}$. From the item perspective, Karako and Manggala \cite{201805} draw on the ideas of the Maximal Marginal Relevance (MMR) re-ranking algorithm, and they define the fairness score based on item embeddings, which measures how the new item $i$ contributes to the embedding difference between two groups. In addition, considering different diversity tolerance of users, Liu et al. \cite{201908} propose a personalized re-ranking method for item fairness. The method of Liu et al. \cite{201908} is only available for a single attribute. Based on it, Sonboli et al. \cite{202005} find that user tolerance is different across item attributes. They define the personalized fairness score based on multiple item attributes and achieve a better trade-off between fairness and utility.

\smallskip
In the \textbf{dynamic ranking} scenario, Morik et al. \cite{202014} propose a re-ranking method based on proportional controller. This method also use a linear strategy to combine recommendation performance and fairness. And they theoretically prove that fairness can be guaranteed when the number of rankings is large enough.
\begin{equation}
\sigma_{\tau} = \mathop{argmax}\limits_{d \in D} (\hat{R}(d|x) + \lambda err_\tau(d)) 
\end{equation}
Here $\hat{R}(d|x)$ is the estimate relevance for item $d$ to user $x$ and $err_{\tau}(d)$ is the error term measuring how fairness will be violated if $d$ is recommended. 

Further, in the dynamic ranking scenario, Yang and Ai \cite{202110} take the marginal fairness into account, i.e., the gain in fairness each time a new item is selected to be added to the recommendation list. They find that the group that maximizes marginal fairness has the lowest current utility-merit ratio. Based on this finding, they propose a probabilistic re-ranking method that jointly optimizes utility and fairness. Specifically, the method will recommend the most relevant item $\tilde{d}_\tau^k$ in a probability of $\lambda$ and the fairness-aware item $\overline{d}_\tau^k$ in a probability of $(1 - \lambda)$.
\begin{equation}
d_\tau^k \sim (\lambda \tilde{d}_\tau^k + (1 - \lambda) \overline{d}_\tau^k) 
\end{equation}
Here $d^k_\tau$ is the item selected for $k^{th}$ position of the presented list at time step $t$, and $\lambda$ is a hyperparameter.

\smallskip
In summary, the slot-wise methods re-rank independently for each user and add items to the re-ranked list one by one. Compared to other re-ranking methods, slot-wise re-ranking tends to be more intuitive and efficient but shortsighted, which may lead to suboptimal performance.

\subsubsection{User-wise}
Apart from picking items slot by slot, we can also directly find the recommendation list for a user based on the optimization goal of the whole list. A popular paradigm is \textbf{integer programming}. The basic idea is that we can treat some decisions of the re-ranking as decision variables and impose some constraints so that the re-ranking problem can be transformed into an integer programming problem.

In the \textbf{group recommendation} scenario, Lin et al. \cite{201704} propose an integer programming-based algorithm to ensure user fairness. The integer programming problem can be formalized as the following Eq.\ref{eq:201704} \cite{201704}. The binary decision variables $X_i$ means whether the recommendation set contains the item $i$. The optimization objective consists of a linear combination of the overall recommendation performance $SW(g,I)$ and the fairness utility $F(g,I)$. 
\begin{equation}
\begin{gathered}
 max \quad \lambda \times SW(g,I) + (1 - \lambda) \times F(g,I) \\
 s.t. \sum_i X_i = K, X_i \in \{0,1\} \label{eq:201704} \\
\end{gathered}
\end{equation}
Here $K$ is the length of the recommendation list, and $\lambda$ is a hyperparameter. This problem is NP-hard. They relax $X_i$ to fractional numbers between zero and one to turn it into a convex optimization problem and then select items with the greatest values of $X_i$.

In \textbf{general recommendation} scenario, Biega et al. \cite{201808} also formalize the fairness problem as an Integer Linear Programming (ILP) problem. The binary decision variables $X_{i,j}$ means whether the i-th item is placed at the j-th position, and the optimization objective is an amortized unfairness metric calculated through the previous ranking results. Unlike the previous work, they prevent large losses in recommended performance by adding constraints related to recommended performance. The ILP problem is formalized as the following Eq.(\ref{eq:201808}) \cite{201808} and then solved by Gurobi, an efficient heuristic algorithm.
\begin{equation}
\begin{gathered}
 minimize \sum_{i = 1}^n \sum_{j = 1}^n |A_i^{l - 1} + w_j - (R_i^{l - 1} + r_i^l)| \times X_{i,j} \\
  s.t. \sum_{j = 1}^k \sum_{i = 1}^n \frac{2^{r_i^l - 1}}{log_2(j + 1)}X_{i,j} \ge \theta \cdot IDCG@k \\
  X_{i,j} \in \{0, 1\}, \sum_i X_{i,j} = 1, \sum_j X_{i,j} = 1 \label{eq:201808}
\end{gathered}
\end{equation}
where $A_i^{l-1}$ denotes the cumulative attention value of the ith item in the previous $l-1$ ranking results, and $w_j$ denotes the attention value assigned to the jth position. $R_i^{l-1}$ denotes the cumulative relevance value, and $r_i^l$ denotes the relevance of the ith item in the current ranking. $\theta$ is the threshold, which means the changed NDCG is required not to fall below a certain value.

Different from the previous work, Singh and Joachims \cite{201819} formalize the problem as a linear programming problem and solve it from a probabilistic ranking perspective. The problem is formalized as the following Eq.(\ref{eq:201819}) \cite{201819}. After the linear programming problem is solved, the final ranking can be sampled through Birkhoff-von Neumann decomposition.
\begin{equation}
\begin{gathered}
P = \mathop{\arg \max}_{P} u^TPv \\
 s.t. \mathbf{1}^T P = \mathbf{1}^T,P\mathbf{1} = \mathbf{1},0 \le P_{i,j} \le 1,\text{$P$ is fair} \label{eq:201819} \\
\end{gathered}
\end{equation}
Here the decision variable $P_{i,j}$ is fractional, which denotes the probability of the item $i$ being placed in the position $j$. The optimization objective $u^TPv$ is the expected recommendation performance, $u_i$ is the predicted relevance score for item $i$ and $v_j$ is the position coefficient for position $j$. Noting that the outliers in rankings may influence the exposure of items, \cite{202205} further extends this approach to mitigate outlierness for fair rankings.

In addition to programming-based methods, Mehrotra et al. \cite{201810} propose several fairness-aware recommendation strategies. The traditional recommendation strategy will maximize the relevance. Assuming that $S$ is the set of candidate recommendation lists, the traditional recommendation strategy can be formalized as $s_u^* = argmax_{s\in S} \phi(u,s)$, where $\phi(u,s)$ is the relevance estimate function, while a recommendation strategy that considers only fairness is $s_u^* = argmax_{s \in S_u} \psi(s)$, where $\psi(s)$ is the fairness estimate function. To combine fairness with relevance, they propose an interpolation strategy, $s_u^* = argmax_{s \in S_u} (1 - \beta) \phi(u,s)+ \beta \psi(s)$ , and a probabilistic strategy as the following Eq.(\ref{eq:201810}) \cite{201810}. 
\begin{equation}
    s_u^*=\begin{cases}
        argmax_{s \in S_u} \psi(s) & if p < \beta\\
        argmax_{s\in S} \phi(u,s) & otherwise \label{eq:201810}
    \end{cases} 
\end{equation}
Here $\beta$ is a hyperparameter.

\smallskip
In summary, the user-wise methods also re-rank independently for each user, and they try to find the optimal list based on the optimization goal of the whole list. Integer programming based on heuristic algorithms is the mainstream method. Compared to slot-wise methods, it considers the information of the whole list to get better performance but is more time-consuming. Besides, compared with the global-wise methods, user-wise methods independently re-rank for each user, which sometimes results in suboptimal performance.

\subsubsection{Global-wise}
Unlike slot-wise and user-wise methods that re-rank a single recommendation list each time, global-wise methods consider global effects and re-rank multiple lists each time, which are more suitable for solving user fairness problems than the user-wise methods.

\smallskip
\textbf{Mathematical programming} is still a common paradigm. Unlike user-wise methods, the decision variable in global-wise re-ranking methods is usually a binary variable indicating whether an item is recommended to a user. We introduce some representative methods below. 

Similar to the user-wise approach, Li et al. \cite{202112} propose an integer programming-based approach to solve the user unfairness problem in the general recommendation scenario, which formalizes the problem as the following Eq.(\ref{eq:202112}) \cite{202112}. The programming problem is solved by Gurobi.
\begin{equation}
\begin{gathered}
\max\limits_{W_{ij}} \sum_{i = 1}^n \sum_{j = 1}^N W_{ij}S_{i,j} \\
 s.t. UGF(Z_1,Z_2,W) < \epsilon, \sum_{j = 1}^N W_{ij} = K, W_{ij} \in \{ 0,1 \} \label{eq:202112}
\end{gathered}
\end{equation}
Here the decision variable $W_{ij}$ is the binary variable indicating whether item $j$ is recommended to the user $i$. $S_{i,j}$ is the preference of user $i$ to item $j$. $Z_1$ and $Z_2$ are two groups of users. $UGF$ is the measurement for user unfairness so that the user group fairness can be guaranteed. $K$ is the length of the recommendation list. $\epsilon$ is a hyperparameter.

While previous work focuses on the fairness of the recommendation performance across different users, Fu et al. \cite{202003} use integer programming to solve fairness problems in the knowledge-based explainable recommendation. The integer programming problem is similar to Li et al. \cite{202112}, and they add a fairness constraint to the optimization problem, which controls the unfairness of explanation diversity in the knowledge graphs.

The above studies focus on user fairness. For item fairness, Sürer et al. \cite{201817} also propose an integer programming-based method. They first formalize the fairness problem as a 0-1 integer programming problem with provider fairness constraints, then relax the conditions using the Lagrangian method, and finally optimize the problem using the subgradient method.

\smallskip
In addition to programming-based methods, there are also some other re-ranking methods. Mansoury et al. \cite{202004,202121} propose a post-processing method for item fairness based on maximum flow matching. The algorithm will build a bipartite graph where the weight between user $u$ and item $i$ is calculated based on the preference of $u$ to $i$ and the degree of $i$, and then iteratively solve the maximum flow matching problem on the graph. Finally, recommendation lists will be constructed based on the candidates identified by the algorithm. Besides, Zhu et al. \cite{202117} propose a parametric post-processing framework for solving the item fairness problem in cold-start scenarios. The method applies an auto-encoder to transform the predicted user preference vector. The transformation needs to satisfy two requirements: the predicted score distribution of under-served items should be as close to the distribution of best-served items as possible, and the predicted score for every user should conform to the same distribution. They propose a generative method and a score scaling method to achieve these requirements.

\smallskip
The above work only considers one-sided fairness. In order to improve \textbf{joint fairness}, Patro et al. \cite{202009} propose a re-ranking method, which consists of two phases. The first phase greedily assigns the most relevant feasible item to each user's recommendation list with limited exposure to each item in the round-robin manner, which ensures that the exposure of each item is greater than a certain value. The second phase does not limit the exposure to the item and recommends the most relevant item for users who have not received enough recommendations. They theoretically prove that this method can guarantee both envy-free fairness for users and maximin-shared fairness for items. Also, at the individual level, Virginie et al. \cite{202122} consider Rawlsian maximin fairness for both users and items. They propose to re-rank by maximizing the concave welfare functions of users and items and provide a tractable inference method based on the Frank-Wolfe algorithm.

While the above methods guarantee joint fairness at the individual level, Wu et al. \cite{202114} propose an offline re-ranking method and an online method that improve item fairness at the group level and user fairness at the individual level. We introduce the algorithm for the offline version here, as the online version is similar. The algorithm will recommend items for all users from position 1 to position k, i.e., the algorithm will not recommend for a certain position until the positions before it has been recommended. The users are sorted by current recommendation quality (random for the first position). Then the algorithm greedily assigns the most relevant feasible item to each user's recommendation list with limited exposure to each provider. If there is no available item, the position will be skipped. After items in position k are selected, the skipped positions will be recommended with an item with the lowest provider exposure to reduce unfairness further. Experiments show that this algorithm can achieve better fairness than the above algorithm of Patro et al. \cite{202009}.

\smallskip
In summary, global-wise methods take global effects into account and re-rank multiple lists each time. Since it re-ranks different users simultaneously, it is more suitable for user fairness than other re-ranking methods and tends to achieve better performance in amortized fairness. However, the dependency between different lists makes the re-ranking process difficult to parallelize and more time-consuming.

\section{Datasets for Fairness Recommendation Study}
\subsection{Overview of Fairness Recommendation Datasets}

As mentioned in Section 2, most work is aimed at improving group fairness in the recommendation. Group fairness requires certain criteria to divide groups, usually the attributes contained in the dataset, such as the gender of users. However, not all recommendation datasets have such attribute information, and existing researchers have not paid the same attention to different attributes. For researchers to easily find fairness-related attributes and the relevant datasets, we survey the recommendation datasets used in the previous fairness studies and list the attributes that researchers have considered in their studies. The reviewed datasets are summarized in Table \ref{tab:dataset}. 

It is worth mentioning that fairness can also be studied on datasets without attributes. If researchers want to study fairness on attribute-free datasets, there are two options to our knowledge. One is to research fairness issues not requiring additional attributes to divide groups, such as the Rawlsian maximin fairness at the individual level. The other is to manually construct attributes based on interaction information, such as item popularity and user activity. These attribute-free datasets used for fairness studies generally only need to contain user and item ID information and ID-aligned user feedback(e.g., rating, click, and purchase) and may not be limited to the datasets summarized below.

{
\begin{table}[h]
\small
\caption{A lookup table for datasets used in existing fairness research in recommendation. We only list the attributes considered in the previous fairness work as fairness-related attributes in the table, and there may be other attributes in the dataset. "-" represents empty. Datasets are arranged in dictionary order.}
\label{tab:dataset}
\resizebox{\width}{!}{%
\begin{minipage}{\columnwidth}
\begin{center}
\begin{tabular}{l|l|l|l|l|l|l} 
\toprule \toprule
\bigcell{l}{\textbf{Datasets}}      & \bigcell{l}{\textbf{Fairness-related} \\ \textbf{User Attributes}}    & \bigcell{l}{\textbf{Fairness-related} \\ \textbf{Item Attributes}}                & \textbf{Users} & \textbf{Items} & \textbf{Interact.} & \textbf{Related Work}                                                   \\ 
\midrule \midrule
\multicolumn{7}{l}{\textbf{with fairness-related attributes:}}
\\
\hline
Amazon         & activity*, gender                  & \bigcell{l}{categories, \\ gender of model}                   & 20.9M & 5.9M & 143.6M & \bigcell{l}{\cite{202105,202112,202114,201901,202003,202017} \\ \cite{202008,201815,202206,202203}}                                                          \\
\hline
Ciao   & -                           & popularity*                                 & 12.3K  & 106K  & 484K      & \cite{202206}                                                                                                           \\ 
\hline
Ctrip Flight   & -                           & airline                                 & 3.8K  & 6K  & 25.1K      & \cite{202114}                                                                                                           \\ 
\hline
Flixter        & -                           & popularity*                            &  1M & 49K  & 8.2M         & \cite{201814}                                                                                                           \\ 
\hline
Google Local   & -                           & business                             & 4.5M & 3.1M & 11.4M         & \cite{202114,202009}                                                                                                    \\
\hline
Insurance      & -                           & \bigcell{l}{ gender, marital status, \\ occupation}       & 1.2K & 21 & 5.3K     & \cite{202115}                                                                                                           \\ 
\hline
Last.FM 1K     & gender, age                       & -                             & 992 & 177K & 904.6K          & \cite{201813}                                                                                                           \\ 
\hline
Last.FM 360K   & gender, age                       & -                                       & 359.3K  & 160.1K & 17.5M & \cite{202109,202009,201813,202121}                                                                                             \\
\hline
ModCloth       & body shape                        & product size                         & 44.7K & 1K & 99.8K         & \cite{202008}                                                                                                           \\ 
\hline
Movielens 100K & -                           & \bigcell{l}{ popularity*, provider, \\ year of movie}                              & 1K & 1.7K &  100K     & \cite{202102,202011,201806,201811,201817,202204}                                                                                             \\ 
\hline
Movielens 1M   & \bigcell{l}{gender, age, \\ occupation} & genres, popularity*  &  6K  & 3.7K & 1M &                        \bigcell{l}{\cite{202102,202107,202109,202115,202117,202004} \\ \cite{202010,202017,201907,201913,201801,201809} \\ \cite{201813,201814,201702,201704,202121,202206}}  \\ 
\hline
Movielens 20M  & -                           & \bigcell{l}{ product company, \\ genres} & 138K  & 27K & 20M & \cite{202117,202014,202018,201912,201816}                                                                               \\ 
\hline
Sushi          & gender, age                       & seafood or not                            & 5K & 100 &  50K    & \cite{201814}                                                                                                           \\ 
\hline
Xing           & premium/standard                           & \bigcell{l}{membership, \\ education degree, \\ working country} & 1.4M & 1.3M & 8.1M &    \cite{202117,201901,202201}                                                                                                    \\ 
\hline
Yelp           & -                           & food genres                    & 2.1M & 160.5K & 8.6M               & \cite{202017,201902,201703}                                                                                             \\ 
\midrule \midrule

\multicolumn{7}{l}{\textbf{without fairness-related attributes:}}                                                                                                           \\ 
\hline
BeerAdvocate   & -                           & -                        & 3.7K & 37.5K &    393K                & \cite{202105}                                                                                                           \\ 
\hline
CiteULike      & -                           & -                        & 5.5K & 16.9K &    204.9K                & \cite{202117} 
                                                                                                   \\ 
\hline
Epinions       & -                           & -                        & 16.5K & 129.3K &  512.7K                  & \cite{202004,202013}                                                                                                          \\ 
\hline
KGRec-music    & -                           & -                        & 5.1K & 8.6K & 751.5K                   & \cite{202010}                                                                                                           \\ 
\hline
Million Song   & -                           & -                       & 1.2M & 380K & 48M                    & \cite{201912}                                                                                                           \\
\hline
Netflix        & -                           & -                        & 480.1K & 17.7K &   100M                 & \cite{201912}                                                                                                           \\ 
\bottomrule \bottomrule
\end{tabular}
\end{center}
\bigskip
$^*$ User activity and item popularity are not attributes in common sense, but researchers also use them to divide groups as attributes. Thus we add them to the table.
\end{minipage}
}
\end{table}
}

The existing datasets for fairness recommendation studies are relatively rich. As seen in Table \ref{tab:dataset}, there are a relatively large number of recommendation datasets containing attribute information. The scenarios of these datasets are diverse, containing movie recommendations (e.g., \textit{Movielens}, \textit{Flixter}, and \textit{Netflix}), e-commerce recommendations (e.g., \textit{Amazon}, \textit{ModCloth}), and job recommendations (e.g., \textit{Xing}).
These datasets contain both large-scale datasets (e.g., \textit{Amazon}) and small-scale datasets (e.g., \textit{Movielens 100K}). The types of interactions in existing datasets are diverse, containing impressions, clicks, and ratings. Moreover, some datasets contain multi-modal information, such as \textit{Amazon} and \textit{Yelp}.

As there is different available information in different scenarios, the attributes considered by researchers are often dataset-specific and vary significantly from one dataset to another, especially for item attributes. For user attributes, gender and age are frequently considered since these attributes are demanded to be fairly treated by anti-discrimination laws. In contrast, item attributes researchers are concerned about are more diverse and contain categories, publishing years, providers, etc. 

Apart from these data-specific attributes, there are also some generic attributes to divide groups, such as user activity and item popularity, which only depend on interactions and can be obtained in all datasets. Researchers who cannot use sensitive attributes for some reasons (e.g., privacy) could consider using interaction information to construct these generic attributes. However, it should be noted that such generic attributes are often dynamic, i.e., an individual may belong to different groups at different times. For example, a current popular item may be cold in the previous time, which means it belonged to the protected group previously but is in the unprotected group now \cite{202102}.

While the existing datasets for fairness studies are diverse, some scenarios and attributes are still worth exploring. For one thing, fairness research can be conducted on some emerging scenarios, such as the short video recommendation scenario, which contains multiple modalities such as video and text. For another, the existing dataset lacks information on some attributes receiving considerable attention, such as race, which is emphasized in anti-discrimination laws \cite{200034}. New data may need to be collected to facilitate relevant research, but privacy concerns must also be considered.

Since most work is attribute-based, we present the datasets with fairness-related attributes and the datasets without fairness-related attributes in Sections 6.2 and 6.3, respectively.

\subsection{Datasets with Fairness-related Attributes}

\quad \textit{\textbf{Amazon.}} This dataset contains product reviews of various categories from Amazon with user and item profiles, including 142.8 million reviews. For user fairness, previous studies divided user groups based on gender \cite{202008} or user activity \cite{202112,202003}. Gender information is not directly accessible, so some researchers use the interaction with Clothing products to infer gender identities \cite{202008}. The active and inactive users can be grouped based on their number of interactions, total consumption capacity, or maximum consumption level \cite{202112}. For item fairness, previous studies usually divided item groups according to their categories \cite{202017}. A few studies use the gender of the model appearing in product images as a grouping method, which is detected through industrial face detection API \cite{202008}.

\textit{\textbf{Ciao.}} This dataset is collected from a popular Web review site of products, which contains user trust networks and ratings. The whole dataset contains 484K ratings from 12.3K users on 106K items. Some researchers use the item popularity to divide item groups \cite{202204}.

\textit{\textbf{Ctrip Flight.}} This dataset contains ticket orders on an international flight route from Ctrip with basic information on customers and some ticket information. The entire dataset includes 3.8K customers, 6K kinds of air tickets and
25K orders. Some researchers treat the airline that the ticket belongs to as the provider, dividing item groups by providers \cite{202114}.

\textit{\textbf{Flixter.}} This dataset is a classical movie recommendation dataset and contains 9.1 million movie ratings from Flixter. Some researchers use the item popularity to divide item groups \cite{201814}. The movies are first sorted by the interaction number in descending order, then the protected and unprotected groups can be divided according to whether the movie is in the top 1\% of the sorted list.

\textit{\textbf{Google Local.}} This dataset is a location recommendation dataset and contains 11.4 million reviews about 3.1 million local businesses from Google Maps. Some researchers divide item groups based on the business of the reviewed item \cite{202114,202009}.

\textit{\textbf{Insurance.}} This dataset is an insurance recommendation dataset in Kaggle, which contains users' information such as gender and occupation. Some researchers divide user groups according to their gender, marital status, and occupation \cite{202115}.

\textit{\textbf{Last.FM 1K.}} This dataset is a music recommendation dataset containing 1K play records of 992 users from Last.FM. It contains user demographic information such as gender and age, which are used to divide user groups by some researchers \cite{201813}.

\textit{\textbf{Last.FM 360K.}} This dataset is similar to Last.FM 1K but has a larger size, including 17 million play records of 360K users. It also contains the gender and age of users, and some researchers divide user groups based on these attributes \cite{202109,201813}.

\textit{\textbf{ModCloth.}} This dataset is an e-commerce recommendation dataset where many products include two human models with different body shapes. The entire dataset contains 100K reviews about 1K clothing products from 44K users. Additionally, there are records of the product sizes which users purchase. For users, some researchers divide users into different body shape groups according to the average size of their purchase \cite{202008}. For items, they divide items into different groups according to the body shape of their models \cite{202008}.

\textit{\textbf{Movielens 100K.}} This dataset is a classical movie recommendation dataset containing 100K movie ratings with user and item profiles. Some researchers divide items into two groups by item popularity, i.e., the number of exposures for each item \cite{202102}. Besides, some studies also divide items into old movies and new movies according to the year of the movie \cite{201811}, and some researchers randomly assign movies among some providers \cite{202011,201817}.

\textit{\textbf{Movielens 1M.}} This dataset is similar to Movielens 100K and has a larger size, including 1 million ratings from 6K users on 4K movies. For users, previous studies divide user groups by their gender, age, and occupation \cite{202107,202109,202115}. For items, the movie genres are seen as a fairness-related attribute \cite{202017}, and item popularity is also considered \cite{202102}.

\textit{\textbf{Movielens 20M.}} This dataset is also collected from Movielens, containing 20 million ratings from 138K users on 27K movies. Some studies consider genres \cite{201816} and production companies of movies \cite{202014} as fairness-related attributes.

\textit{\textbf{Sushi.}} This dataset includes 5K responses to a questionnaire survey of preference in sushi which contains preference data and demographic data. Some researchers consider three types of fairness-related attributes: age, gender, and whether or not a type of sushi is seafood \cite{201814}.

\textit{\textbf{Xing.}} This dataset is a user-view-job dataset, which contains 320M of interactions with user and item profiles such as career level. Some researchers \cite{201901} consider the membership, education degree, and working country as fairness-related attributes for items. In addition, whether the user is premium or not is also regarded as a fairness-related attribute for users \cite{202201}.

\textit{\textbf{Yelp.}} This dataset is a business review dataset. Some studies only focus on the restaurant business and divide item groups based on the food genres of restaurants \cite{202017}.

\subsection{Datasets without Fairness-related Attributes}
We also survey the recommendation datasets without fairness-related attributes in current fairness studies, which are usually used in research on individual fairness. 

\textit{\textbf{BeerAdvocate.}} This dataset \cite{202105} contains 1.5 million beer reviews from the BeerAdvocate, including products, user information, and their ratings. Some researchers \cite{202105} leverage the reviews in this dataset to enhance the representation of non-mainstream users by adding a textual information reconstruction task.

\textit{\textbf{CiteULike.}} This dataset \cite{202117} includes about 200K records of user preferences toward scientific articles from 5K users. Some researchers \cite{202117} utilize this dataset to explore Rawlsian maximin fairness issues in the cold start scenario.

\textit{\textbf{Epinions.}} This dataset \cite{202004,202013} is collected from a Web review site of products, which contains user bidirectional connections and ratings. The whole dataset contains 512K ratings from 16K users on 129K items. Some researchers \cite{202004,202013} use this dataset to explore consistent fairness issues at the individual level.

\textit{\textbf{KGRec-music.}} This dataset \cite{202010} is a music recommendation dataset that contains knowledge graphs. The dataset includes about 750K interactions from about 5K users on 8K songs. Some researchers use this dataset \cite{202010} to investigate the individual fairness of users in group recommendations.

\textit{\textbf{Million Song.}} This dataset \cite{201912} contains audio attributes and metadata for a million tracks from contemporary popular music, including 1 million songs with 515K dated tracks. Some researchers use this dataset \cite{201912} to explore the individual fairness of items.

\textit{\textbf{Netflix.}} This dataset \cite{201912} is a movie recommendation dataset from Netflix, containing 100 million ratings from 480K users over 17K movies. Some researchers \cite{201912} leverage this dataset to study the individual fairness of items.

\section{Future Directions}
Fairness is essential for recommender systems and needs to be further exploited. In this section, we discuss some promising future directions for fairness in the recommendation from the perspectives of definition, evaluation, algorithm design, and explanation.

\subsection{Definition}
\quad \textit{\textbf{A general deﬁnition of fairness.}}
As mentioned earlier, many different definitions of fairness have been applied in recommender systems. These fairness definitions may conflict with each other. For example, calibrated fairness may be damaged when ensuring Rawlsian maximin fairness and vice versa. Therefore, it is important to determine the priority between different definitions of fairness, but to our knowledge, there is no work on it yet. In addition, a general definition of fairness may not exist. The appropriate fairness definition may vary in different scenarios. A consensus in each scenario would be helpful.

\subsection{Evaluation}
\quad \textit{\textbf{Fair comparison between different fairness methods.}}
No effective benchmarks may result in non-reproducible evaluation and unfair comparison and damage the development of the research community. Existing fairness research suffers from this problem since many different fairness measurements and data-processing strategies exist. Hence, it is necessary to propose a standard experimental setting including but not limited to data preprocessing methods, hyper-parameter tuning strategies, and evaluation metrics.

\textit{\textbf{Dataset for new emerging scenarios.}}
Existing datasets for fair recommendation studies are diverse, but there is a lack of investigation on some emerging recommendation scenarios. For example, short video recommendation plays an important role today, and it contains multiple modal information, which is quite different from traditional recommendation scenarios. However, there is a lack of fairness-related work on short-video recommendation datasets. Whether there are serious unfairness issues in these emerging scenarios and how to address them deserve to be explored.

\subsection{Algorithm Design}
\quad \textit{\textbf{A win-win for fairness and accuracy.}}
Existing methods often improve the fairness in recommendation with a loss in recommendation performance, and many papers have revealed such a tradeoff between fairness and performance. In the optimal case, fairness is not always in conflict with recommendation performance; for example, for user fairness, both recommendation performance and fairness are optimal if all users receive the most accurate recommendation list. In practice, some work on classification tasks has also found that improving fairness may improve overall accuracy \cite{200047}. For industrial recommender systems, the degradation of recommendation performance may result in an unacceptably large loss of revenue. For successfully applying fairness methods to recommender systems, it is necessary to investigate methods to improve fairness while ensuring recommendation accuracy.

\textit{\textbf{Fairness for both user and item.}}
Many approaches to improving fairness have been proposed, but most focus on only one type of user fairness or item fairness. However, both user and item fairness are essential and should be guaranteed in most recommender systems. Hence, it is worthwhile to propose adequate methods for joint fairness. Note that there may be a natural conflict between user fairness and item fairness\cite{202114}, which also makes joint fairness a challenging topic.

Joint fairness issues can be addressed using different types of methods. First, current practices about data-oriented methods only consider one-sided fairness, and it is worth exploring how to adapt for joint fairness. Second, the joint fairness problem can be regarded as multi-objective learning. The trade-off between multiple fairness and recommendation accuracy might be improved by drawing on Pareto optimization \cite{200059} and "seesaw phenomenon" related work such as \cite{200060}. Third, most existing re-ranking methods for joint fairness are non-parametric re-ranking algorithms. It is worth investigating how to design learnable re-ranking algorithms, as they have shown better performance on one-sided fairness \cite{202117}.

\textit{\textbf{Fairness beyond accuracy.}}
Most user fairness work focuses on one evaluation criterion, i.e., accuracy. However, there are many other measurements beyond accuracy, such as diversity, unexpectedness, and serendipity, which are also closely related to user satisfaction. Research has found that unfairness also exists in these measurements \cite{202119}. Therefore, we also need to consider more measurements beyond accuracy when ensuring user fairness.

\textit{\textbf{Causal inference for fairness.}}
Eliminating unfairness at the causal level is considered essential and has received a growing interest in machine learning \cite{200057,200058}. Similarly, causal inference in recommendation has attracted increasing attention and has become a popular recommendation debiasing technique \cite{200055,200056}. However, as we mentioned in Section 2, fairness and bias are different. Only a little work \cite{202115} in recommendation focuses on fairness at the causal level. In our opinion, two problems need to be solved. The first problem is how to construct causal graphs for fair recommendations. Most current work focuses on the models based on only ID information. For these models, we can manually design causal graphs. However, for models using additional features, since there exists some causal relationship between these features as well, how to construct causal graphs becomes a challenging problem. The second problem is how to eliminate the influence of unfair factors based on the already constructed causal graphs, especially in the complex causal graphs mentioned above. There is still much room for using causal inference to achieve fair recommendations.

\textit{\textbf{Fairness with missing data.}}
Existing studies usually assume all fairness-related attributes are available in the dataset. However, there are users or items whose fairness-related attributes are missing in many real-world scenarios. For example, some users may fill in their gender as confidential or even false information. In this case, we cannot identify whether the sensitive group is treated unfairly, and existing fairness methods will be ineffective. Therefore, it is necessary to investigate methods to improve fairness when fairness-related information is missing. Solving this problem also helps to reduce the risk of sensitive information leakage since we only depend on partial information. There has been some related work on classification tasks \cite{200030,200031,200016,202108,200019}, while in recommender systems, it is still a problem to be explored.

\textit{\textbf{Fairness in a real system.}}
Industrial recommender systems usually consist of three phases: recall, ranking, and re-ranking. Some objectives other than accuracy, such as diversity, are often involved in the re-ranking phase. Existing studies have found that some post-processing techniques, such as diversity re-ranking, may increase user unfairness \cite{201816}, implying that some fairness methods applied before re-ranking will be ineffective. Therefore, it is necessary to investigate how to improve fairness more effectively in real-world systems. For example, we can consider adding fairness-oriented recall in the recall phase.
On the other hand, industrial recommender systems usually require a short response time, so more efficient re-ranking algorithms need to be proposed. Moreover, industrial recommender systems often have multiple objectives, such as reading time and purchase. However, the fairness of the corresponding model for each goal cannot guarantee the fairness of the final recommendations \cite{202101}, which also poses a challenge for applying fairness in industrial systems.

\subsection{Explanation}
\quad \textit{\textbf{What are the causes of unfairness?}}
Explainability is crucial for recommender systems as it can improve the persuasiveness of recommendations, increase user satisfaction, and enhance the transparency of the whole system \cite{200048}. Explaining why unfairness occurs can deepen the understanding of unfairness and facilitate the design of more effective fairness methods. Although many approaches have been developed to improve fairness in the recommendation, there is relatively little work on the explainability of fairness, i.e., why unfairness occurs. Only a few studies \cite{202104} have theoretically proven that a specific class of models leads to unfairness in recommendation results. Causal inference-based and more theoretical analyses remain a challenge.

\section{Conclusion}
Unfairness is widespread in recommender systems, which has attracted increasing attention in recent years, and a series of fairness definitions, measurements, and methods have been proposed. This survey systematically reviews fairness-related research in the recommendation and summarizes current fairness work from multiple perspectives, including definitions, views, measurements, datasets, and methods.

For fairness definitions, previous studies mainly focus on outcome fairness, which we further classify according to different targets and concepts. We find that group fairness is the most common target, and consistent fairness and calibrated fairness are the most common concepts. As for fairness views, we present some views to classify fairness issues in the recommendation, containing fairness subjects, fairness granularity, and fairness optimization objects.
For fairness measurements, we introduce representative measurements of existing work and summarize common metrics of different fairness definitions. As for fairness methods, we review representative studies from data-oriented methods, ranking methods, and re-ranking methods. It is common for researchers to develop ranking and re-ranking methods to achieve fair recommendations, while there are only a few studies on adjusting data to improve fairness. Additionally, we summarize fairness-related recommendation datasets occurring in previous fairness work for researchers to find relevant datasets easily.

Furthermore, we discuss some promising future research directions from different perspectives for fairness in the recommendation.
In terms of definitions, which definition of fairness is most proper to recommender systems may be an important problem for future work. As for evaluation, we could develop effective benchmarks to compare different fairness methods fairly. In terms of algorithm design, we discuss some promising future work containing fairness methods for both user and item and fairness methods beyond accuracy. In terms of explanation, explaining why unfairness exists could be a problem worth exploring.
Finally, we hope this survey may help readers better understand fairness issues in recommender systems and provide some inspiration.


\bibliographystyle{ACM-Reference-Format}
\bibliography{main}

\end{document}